\documentclass[twocolumn]{aastex631}

\usepackage{array}
\usepackage{makecell}
\usepackage{hyperref}
\usepackage{multirow}
\usepackage{makecell}
\usepackage{rotating}

\usepackage{amsmath}

\newcommand{\ebvgas}{$E(B-V)_{\text{gas}}$}

\newcommand{\one}{~\textsc{i}}
\newcommand{\ii}{~\textsc{ii}}
\newcommand{\iii}{~\textsc{iii}}

\newcommand{\fint}{$f_\mathrm{int}$}
\newcommand{\fobs}{$f_\mathrm{obs}$}

\newcommand{\cgs}{~\mathrm{erg}~\mathrm{s}^{-1}~\mathrm{cm}^{-2}}

\shorttitle{Nebular Attenuation Curve at $z=4.41$}
\shortauthors{Sanders et al.}

\begin{document}

\title{The AURORA Survey: The Nebular Attenuation Curve of a Galaxy at $z=4.41$ from Ultraviolet to Near-Infrared Wavelengths}

\author[0000-0003-4792-9119]{Ryan L. Sanders}\affiliation{Department of Physics and Astronomy, University of Kentucky, 505 Rose Street, Lexington, KY 40506, USA}

\email{email: ryan.sanders@uky.edu}

\author[0000-0003-3509-4855]{Alice E. Shapley}\affiliation{Department of Physics \& Astronomy, University of California, Los Angeles, 430 Portola Plaza, Los Angeles, CA 90095, USA}

\author[0000-0001-8426-1141]{Michael W. Topping}\affiliation{Steward Observatory, University of Arizona, 933 N Cherry Avenue, Tucson, AZ 85721, USA}

\author[0000-0001-9687-4973]{Naveen A. Reddy}\affiliation{Department of Physics \& Astronomy, University of California, Riverside, 900 University Avenue, Riverside, CA 92521, USA}

\author[0000-0002-4153-053X]{Danielle A. Berg}\affiliation{Department of Astronomy, The University of Texas at Austin, 2515 Speedway, Stop C1400, Austin, TX 78712, USA}

\author[0000-0002-4989-2471]{Rychard J. Bouwens}\affiliation{Leiden Observatory, Leiden University, NL-2300 RA Leiden, Netherlands}

\author[0000-0003-2680-005X]{Gabriel Brammer}\affiliation{Niels Bohr Institute, University of Copenhagen, Lyngbyvej 2, DK2100 Copenhagen \O, Denmark}

\author[0000-0002-1482-5818]{Adam C. Carnall}\affiliation{Institute for Astronomy, University of Edinburgh, Royal Observatory, Edinburgh, EH9 3HJ, UK}

\author[0000-0002-3736-476X]{Fergus Cullen}\affiliation{Institute for Astronomy, University of Edinburgh, Royal Observatory, Edinburgh, EH9 3HJ, UK}

\author[0000-0003-2842-9434]{Romeel Dav\'e}\affiliation{Institute for Astronomy, University of Edinburgh, Royal Observatory, Edinburgh, EH9 3HJ, UK}

\author{James S. Dunlop}\affiliation{Institute for Astronomy, University of Edinburgh, Royal Observatory, Edinburgh, EH9 3HJ, UK}

\author[0000-0001-7782-7071]{Richard S. Ellis}\affiliation{Department of Physics \& Astronomy, University College London. Gower St., London WC1E 6BT, UK}

\author[0000-0003-4264-3381]{N. M. F\"orster Schreiber}\affiliation{Max-Planck-Institut f\"ur extraterrestrische Physik (MPE), Giessenbachstr.1, D-85748 Garching, Germany}

\author[0000-0002-0658-1243]{Steven R. Furlanetto}\affiliation{Department of Physics \& Astronomy, University of California, Los Angeles, 430 Portola Plaza, Los Angeles, CA 90095, USA}

\author[0000-0002-3254-9044]{Karl Glazebrook}\affiliation{Centre for Astrophysics and Supercomputing, Swinburne University of Technology, P.O. Box 218, Hawthorn, VIC 3122, Australia}

\author[0000-0002-8096-2837]{Garth D. Illingworth}\affiliation{Department of Astronomy and Astrophysics, UCO/Lick Observatory, University of California, Santa Cruz, CA 95064, USA}

\author[0000-0001-5860-3419]{Tucker Jones}\affiliation{Department of Physics and Astronomy, University of California Davis, 1 Shields Avenue, Davis, CA 95616, USA}

\author[0000-0002-7613-9872]{Mariska Kriek}\affiliation{Leiden Observatory, Leiden University, NL-2300 RA Leiden, Netherlands}

\author[0000-0003-4368-3326]{Derek J. McLeod}\affiliation{Institute for Astronomy, University of Edinburgh, Royal Observatory, Edinburgh, EH9 3HJ, UK}

\author{Ross J. McLure}\affiliation{Institute for Astronomy, University of Edinburgh, Royal Observatory, Edinburgh, EH9 3HJ, UK}

\author[0000-0002-7064-4309]{Desika Narayanan}\affiliation{Department of Astronomy, University of Florida, 211 Bryant Space Sciences Center, Gainesville, FL, USA}

\author[0000-0001-5851-6649]{Pascal A. Oesch}\affiliation{Department of Astronomy, University of Geneva, Chemin Pegasi 51, 1290 Versoix, Switzerland}\affiliation{Niels Bohr Institute, University of Copenhagen, Lyngbyvej 2, DK2100 Copenhagen \O, Denmark}

\author[0000-0003-4464-4505]{Anthony J. Pahl}\affiliation{The Observatories of the Carnegie Institution for Science, 813 Santa Barbara Street, Pasadena, CA 91101, USA}

\author{Max Pettini}\affiliation{Institute of Astronomy, Madingley Road, Cambridge CB3 OHA, UK}

\author[0000-0001-7144-7182]{Daniel Schaerer}\affiliation{Department of Astronomy, University of Geneva, Chemin Pegasi 51, 1290 Versoix, Switzerland}

\author{Daniel P. Stark}\affiliation{Steward Observatory, University of Arizona, 933 N Cherry Avenue, Tucson, AZ 85721, USA}

\author[0000-0002-4834-7260]{Charles C. Steidel}\affiliation{Cahill Center for Astronomy and Astrophysics, California Institute of Technology, MS 249-17, Pasadena, CA 91125, USA}

\author[0000-0001-5940-338X]{Mengtao Tang}\affiliation{Steward Observatory, University of Arizona, 933 N Cherry Avenue, Tucson, AZ 85721, USA}

\author[0000-0003-1249-6392]{Leonardo Clarke}\affiliation{Department of Physics \& Astronomy, University of California, Los Angeles, 430 Portola Plaza, Los Angeles, CA 90095, USA}

\author[0000-0002-7622-0208]{Callum T. Donnan}\affiliation{Institute for Astronomy, University of Edinburgh, Royal Observatory, Edinburgh, EH9 3HJ, UK}

\author{Emily Kehoe}\affiliation{Department of Physics \& Astronomy, University of California, Los Angeles, 430 Portola Plaza, Los Angeles, CA 90095, USA}

\begin{abstract}
We use {\it JWST}/NIRSpec observations from the Assembly of Ultradeep Rest-optical Observations Revealing Astrophysics (AURORA) survey
 to constrain the shape of the nebular attenuation curve of a star-forming galaxy at $z=4.41$, GOODSN-17940.
We utilize 12 H\one\ recombination lines to derive the attenuation curve spanning
 optical to near-infrared wavelengths ($3751-9550$~\AA).
We then leverage a high-S/N spectroscopic detection of the rest-frame ultraviolet continuum in combination with
 rest-UV photometric measurements to constrain the shape of the curve at ultraviolet wavelengths.
While this UV constraint is predominantly based on stellar emission, the large measured equivalent widths of H$\alpha$ and H$\beta$
 indicate that GOODSN-17940 is dominated by an extremely young stellar population $<10$~Myr in age such that 
 the UV stellar continuum experiences similar attenuation to that of the nebular emission.
The resulting combined nebular attenuation curve spans $1400-9550$~\AA\ and has a shape that deviates significantly from commonly assumed
 dust curves in high-redshift studies.
Relative to the Milky Way, SMC, and Calzetti curves, the new curve has a steeper slope at long wavelengths ($\lambda>5000$~\AA)
 while displaying a similar slope across blue-optical wavelengths ($\lambda=3750-5000$~\AA).
In the ultraviolet, the new curve is shallower than the SMC and Calzetti curves and displays no significant 2175~\AA\ bump.
This work demonstrates that the most commonly assumed dust curves are not appropriate for all high-redshift galaxies.
These results highlight the ability to derive nebular attenuation curves for individual high-redshift sources with deep {\it JWST}/NIRSpec spectroscopy,
 thereby improving the accuracy of physical properties inferred from nebular emission lines.
\end{abstract}

\section{Introduction} \label{sec:intro}

Accurately determining the physical properties of galaxies from observations at rest-frame optical or ultraviolet
 wavelengths requires robust corrections for the impact of dust on the intrinsically emitted spectrum.
The dust attenuation law describes the wavelength-dependent effects of dust integrated over multiple sightlines,
 relevant for observations on sub-galactic spatial scales and those integrated over entire galaxies.
It accounts for both the loss of light along the line of sight due to absorption or scattering by intervening
 dust determined by the dust grain properties (i.e., extinction)
 and also the effects arising from the geometric distribution of stars, gas, and dust inside of the galaxy including
 scattering of light back into the line of sight and variations in optical depth to different regions
 in the galaxy \citep[e.g.,][]{sal2020}.
The presence of dust in the interstellar medium (ISM) of galaxies both attenuates light at any particular wavelength
 and reddens the intrinsic spectrum.
Consequently, galaxy properties inferred from the absolute brightness at optical or ultraviolet wavelengths
 or the relative brightness (i.e., galaxy colors or emission-line ratios) between two widely separated wavelengths
 rely on a robust dust correction.
Such properties include those primarily sensitive to stellar continuum emission such as stellar masses from fitting
 broadband photometry and star-formation rates (SFRs) derived from the rest-frame ultraviolet luminosity,
 but also those derived from nebular emission lines such as SFR from dust-corrected H$\alpha$ luminosity
 and gas-phase metallicity or ionization parameter from rest-optical or rest-UV line ratios.
Despite the geometrically extended nature of nebular emitting regions, it is a common practice to use the
 average Milky Way extinction curve \citep[e.g.,][]{car1989} to correct for dust attenuation effects on nebular emission lines.

The dust attenuation curve has been studied extensively at high redshifts ($z\gtrsim2$) through different methods,
 including ``direct'' derivations of the curve leveraging samples with spectroscopic Balmer decrement
 measurements to sort galaxies according to the degree of attenuation and characterize the variation in SED
 shape as a function of attenuation \citep[e.g.,][]{red2015,shi2020,bat2022} or performing SED fitting while
 varying the assumed attenuation curve \citep[e.g.,][]{bua2012,kri2013,sco2015,sal2016};
 and indirect constraints based on the position of galaxies in the diagram comparing the
 infrared excess (IRX=$L_\mathrm{UV}/L_\mathrm{IR}$) to the UV continuum slope ($\beta$)
 \citep[e.g.,][]{red2010,red2018,bou2016,mcl2018}.
Collectively, these works suggest significant variation in the dust attenuation curve
 among the high-redshift galaxy population, suggesting that a universal dust law does not apply,
 similar to results at $z\sim0$ \cite[e.g.,][]{sal2018}.
This variation correlates with galaxy properties, with younger, low-mass, low-metallicity, and low-attenuation galaxies
 generally displaying curves with UV slopes similar to or steeper than that of the SMC extinction curve, and galaxies that are more massive
 with older stellar populations, higher metallicities, more attenuation, and higher SFRs
 typically showing shallower attenuation curves more similar
 to the \citet{cal2000} starburst curve \citep[e.g.,][]{red2010,kri2013,bou2016,mcl2018,red2018,shi2020,shi2020b,boq2022}.
However, these studies have focused almost exclusively on the stellar attenuation curve.
The dust attenuation curve appropriate for the correction of nebular emission lines that are used to derive
 many fundamental physical properties may be distinct from the stellar curve in shape and normalization \citep[e.g.,][]{sal2020}.

There is currently only one study in the literature that directly constrained the nebular attenuation curve at high redshift from H\one\ emission lines.
\citet{red2020} derived the mean nebular attenuation curve for a sample of $\sim500$ star-forming galaxies
 at $z=1.4-2.6$ from the MOSDEF survey \citep{kri2015}.
Due to the sensitivity limits of the Keck/MOSFIRE spectra, in which only H$\alpha$ and H$\beta$
 were detected for all but the brightest few sources, these authors relied on stacked
 spectra to detect higher-order H\one\ Balmer lines up to H$\epsilon$ at $\lambda_\mathrm{rest}=3971$~\AA.
They derived an average $z\sim2$ nebular attenuation curve that closely matched the shape of the \citet{car1989}
 Milky Way extinction curve over $4000-6600$~\AA.
There are a number of ways in which it is desirable to improve upon the results of \citet{red2020} in light
 of the near-infrared spectroscopic capabilities offered by the James Webb Space Telescope ({\it JWST}), including
 carrying out a similar analysis on individual galaxies and extending the wavelength range with the detection
 of higher-order Balmer lines (importantly reaching $\lambda_\mathrm{rest}\approx3700$~\AA\ for robust [O\ii]$\lambda$3728 dust corrections)
 and rest-frame near-infrared Paschen lines. 

The sensitivity of the NIRSpec instrument \citep{jak2022} onboard {\it JWST} has opened a new era
 of emission line studies of high-redshift sources, providing the ability to move beyond the
 handful of brightest ``strong lines'' and obtain novel physical constraints utilizing weak emission lines
 for individual sources.
One of the analyses made possible by these enhanced capabilities is the derivation of the nebular attenuation curve
 on an individual galaxy basis, improving upon past work that relied on composite spectra of hundreds of galaxies \citep{red2020}.
With suitably deep integrations, NIRSpec's medium- and high-resolution spectroscopic modes in particular
 enable the detection of high-order Balmer and Paschen series H\one\ recombination lines that would
 otherwise be blended in low-resolution prism or grism observations.
Such data can be leveraged to measure a large number of H\one\ lines
 for use in constraining dust attenuation, increasing the wavelength range over which the dust curve can be characterized.
We thus now have the ability to determine dust attenuation curves over wide wavelength ranges for individual high-redshift galaxies.
This advance can lead to an increase in the accuracy of physical properties inferred for each particular target and makes it possible to
 assess the amount of variation of dust curves among the population and how that variation depends on galaxy properties such
 as mass, SFR, and metallicity.

In this paper, we derive the nebular attenuation curve spanning from far-ultraviolet to near-infrared wavelengths
 for one galaxy at $z=4.4115$, GOODSN-17940 (R.A.=12:36:35.483; Dec.=+62:13:50.04).
This source is a starburst galaxy with extremely strong nebular emission lines, a spectrum of which
 was first published in \citet{sha2017} displaying $4-5\sigma$ detections of [O\ii]$\lambda\lambda$3727,3730,
 [Ne\iii]$\lambda$3870, and H$\gamma$ based on Keck/MOSFIRE observations.
{\it JWST}/NIRSpec medium-resolution spectroscopy was recently obtained for GOODSN-17940 as part of the
 Assembly of Ultradeep Rest-optical Observations Revealing Astrophysics (AURORA) Cycle 1 program.
The deep (24~h on-source) and continuous $1-5~\mu$m AURORA observations have yielded a remarkably detailed
 spectrum of GOODSN-17940 spanning rest-frame wavelengths of 1800~\AA\ to 1~$\mu$m, with over 70
 individual emission and absorption features detected with a signal-to-noise ratio $S/N\ge3$.
This work is the first in a series of papers leveraging this extremely rich data set for GOODSN-17940 to obtain a detailed comprehensive
 picture of this galaxy rivaling the information content available for any other single high-redshift source.
We begin this series with an analysis of the nebular dust attenuation curve since this component is required to
 correct measured line strengths for the effects of dust attenuation and reddening for robust derivations of
 properties such as SFR, metallicity, and ionization parameter.

This paper is organized in the following way.
We describe the observations, measurements, and derived properties in Section~\ref{sec:obs}.
In Section~\ref{sec:results}, we present the derivation of the nebular attenuation curve, first at
 optical to near-infrared wavelengths based on H\one\ emission lines (Sec.~\ref{sec:opt}), then at ultraviolet
 wavelengths using the observed rest-UV continuum shape (Sec.~\ref{sec:uv}), and presenting the form of the full
 ultraviolet-to-near-IR attenuation curve in Sec.~\ref{sec:comb} (equation~\ref{eq:ktot}).
In Section~\ref{sec:discussion}, we discuss the implications of the form of the newly derived dust curve relative to commonly-assumed
 curves.
We summarize the results and state our conclusions in Section~\ref{sec:conclusions}.
We adopt the \citet{asp2021} solar abundances (i.e., 12+log(O/H$)_{\odot}=8.69$)
 and a cosmology described by $H_0=70\mbox{ km  s}^{-1}\mbox{ Mpc}^{-1}$, $\Omega_m=0.30$, and $\Omega_{\Lambda}=0.7$.
Emission-line wavelengths are given in the vacuum rest frame.
Color excesses are given in magnitudes.

\section{Observations amd Measurements}\label{sec:obs}

\subsection{The AURORA survey}

The AURORA survey (Program ID: 1914; PIs: A. Shapley \& R. Sanders) utilized the NIRSpec instrument onboard {\it JWST}
 to obtain multi-object spectroscopy of $\sim100$ galaxies at $z\sim2-10$ using the Microshutter Assembly (MSA).
Observations were taken for 2 pointings in CANDELS \citep{gro2011,koe2011} extragalactic legacy fields: one in COSMOS and one in GOODS-N.
In this analysis, we focus on one galaxy in GOODS-N with an ID number of 17940, though we describe some aspects of the data reduction
 and measurements relevant for the full AURORA sample below.
For each pointing, one MSA mask was observed in the medium resolution G140M/F100LP, G235M/F170LP, and G395M/F290LP.
The integration times were tiered based on the relative sensitivity of each grating with an approximate total
 of 24~h on-source per pointing, broken down into 44~ks (12.3~h) in G140M, 29~ks (8.0~h) in G235M, and 15~ks (4.2~h) in G395M.
A mixture of NRSIRS2 and NRSIRS2RAPID readout modes was used to meet data volume limit requirements.
Each MSA slitlet was composed of three microshutters, and a 3-point nod pattern was employed accordingly.
This setup provides continuous $R\sim1000$ coverage over $1-5~\mu$m in wavelength (excepting chip gaps) with
 a nearly uniform limiting line sensitivity across the full wavelength range.
The integration times were chosen to yield a uniform $5\sigma$ limiting line flux of $10^{-18}~\mathrm{erg}~\mathrm{s}^{-1}~\mathrm{cm}^{-2}$
 after slit loss correction based on the NIRSpec Exposure Time Calculator,
 though we found that the on-sky performance is roughly a factor of two better.

The GOODS-N and COSMOS masks were populated with 51 and 46 targets, respectively.
The primary science goal of AURORA is to measure direct-method metallicities of high-redshift galaxies.
Accordingly, the pointings and mask designs were optimized to contain the maximum number of
 targets at $z\sim2-4$ that were expected to yield detections of one or more
 temperature-sensitive auroral emission lines (e.g., [O\iii]$\lambda$4364, [O\ii]$\lambda\lambda$7322,7332).
Potential targets were selected from the rest-optical emission-line catalogs of the MOSDEF survey \citep{kri2015},
 the extreme emission-line galaxy (EELG) catalogs of \citet{tan2019},
 and photometric catalogs of the 3D-{\it HST} survey \citep{ske2014}.
For sources with existing rest-optical emission-line measurements, we used the ratios of the strong lines
 to estimate the high-ionization (i.e., O$^{2+}$) zone electron temperature $T_3$ and the [O\iii]$\lambda$4364 flux based on the relation between
 $T_3$ and strong-line ratios displayed by local H\ii\ regions \citep{san2017}.
Fitted relations between the brightness ratio of auroral lines of [O\ii], [S\iii], [S\ii], and [N\ii] to [O\iii]$\lambda$4364
 based on local H\ii\ regions from the CHAOS survey \citep{ber2020} were then used to estimate the flux of other auroral lines.
This sample was supplemented with 3D{\it HST} sources with [O\iii]$\lambda$5008 flux $>10^{-16}\cgs$ for which we expect to detect
 [O\iii]$\lambda$4364 if $T_3>10,000$~K.
This selection yielded 249 candidate auroral-line targets in COSMOS and 210 candidates in GOODS-N spanning $z=1.2-4.7$.

Auroral targets were added to MSA masks following a scheme described in \citet{sha2024}
 that prioritized targets with brighter predicted auroral line fluxes, multiple auroral lines predicted above the detection threshold,
 and [O\ii]$\lambda\lambda$3727,3730 within the G140M wavelength coverage (i.e., $z\ge1.63$).
The final pointings in each field were chosen to maximize the number of high-priority auroral targets on each mask,
 with 16 on the COSMOS mask and 20 on the GOODS-N mask.
The subject in this analysis, GOODSN-17940, was one of these high-priority auroral targets based on pre-existing MOSFIRE spectroscopy
 and its 3D-{\it HST} photometry.
Additional filler targets were added to both masks as described in \citet{sha2024}.

\subsection{Data Reduction}

Data were reduced using a pipeline that is a combination of the STScI distributed pipeline (version 1.13.4) and custom routines.
Detector bias, dark current, and detector gain were first corrected for.
After correcting the images for linearity, the count rate was determined from the readout groups in each pixel
 and large jumps between subsequent readouts were used to identify and remove cosmic rays or ``snowball'' events.
The \texttt{nsclean} package was used to correct the count rate files for $1/f$ noise \citep{rau2024}.
For each slit, 2D spectra were extracted from the corrected rate files and were flat fielded, wavelength calibrated,
 and flux calibrated.
The two observations at each nod position were compared to exclude remaining artifacts or cosmic rays not removed in earlier steps.
The final 2D spectrum was then obtained by interpolating the cutout 2D spectra at each nod position onto a common wavelength
 grid and combining them following the adopted 3-point dither pattern.

In each grating, 1D spectra were optimally extracted \citep{hor1986} using the spatial profile defined by the 
brightest detected emission-line in that grating or the integrated continuum if no emission lines were present.
If neither lines nor continuum were detected in a grating, then spatial profile was indirectly estimated by averaging the extracted
 profile of all other gratings with detected features, though this case only occurred for 4/288 grating+target combinations
 thanks to the depth of the AURORA spectra.
Wavelength pixels containing cosmic rays, artifacts, or contamination from dithers onto neighboring sources
 were flagged during 1D extraction and masked during further analysis.
For GOODSN-17940, the extracted spatial profile was determined by C\iii]$\lambda$1908 in G140M,
 [O\iii]$\lambda$5008 in G235M, and H$\alpha$ in G395M.

\subsection{Slit Loss Correction}

Wavelength-dependent slit losses were calculated to correct for light falling outside of the microshutters and 1D extraction window,
 accounting for the wavelength-dependent {\it JWST}/NIRSpec PSF, following the method described in \citet{red2023}.
Briefly, after masking out other sources using the segmentation map, the {\it JWST}/NIRCam F115W image of each target was
 convolved with Gaussian kernels with widths determined to match the wavelength variation of the {\it JWST} PSF over the full
 range of wavelenths covered by the AURORA observations, providing a model of the light profile as a function of wavelength.
At each wavelength, the fraction of total light falling inside the 1D extraction window was calculated and the extracted
 1D spectrum and error spectrum were divided by this value to correct for slit losses.

\subsection{Flux Calibration}\label{sec:fluxcal}

The final flux calibration of the AURORA spectra 
 first addressed the relative flux calibration between gratings and then
 set the absolute flux calibration across all gratings.
The three medium-resolution gratings utilized for AURORA observations provide $2000-3000$~\AA\ of overlapping
 observed-frame wavelength coverage between neighboring gratings, enabling an assessment of the relative flux calibration between gratings.
The depth of the AURORA observations enable significant detections of emission lines ($S/N\ge5$) and/or continuum
 (median $S/N$ per pixel $\ge3$) in the overlap regions for the majority of targets.
For each pointing, the significantly-detected emission-line fluxes or integrated continuum flux densities in these overlap regions are compared
 to determine the median offset in flux scale between the G140M and G395M gratings relative to the G235M grating.
The median flux ratios were found to be near unity for all grating pairs except for G395M/G235M in COSMOS.
The median scale factors were 0.96 for G140M/G235M in GOODSN, 1.02 for G395M/G235M in GOODSN,
 0.96 for G140M/G235M in COSMOS, and 0.80 for G395M/G235M in COSMOS.
Limiting the comparison to only line fluxes or only integrated continuum yields similar results.
The G140M and G395M spectra were scaled by these median ratios for all objects lying inside of the intrinsic 1$\sigma$ scatter about the
 medians, where this intrinsic scatter was 8\%\ on average between the pointing+grating combinations.
Outliers lying farther from the median were scaled such that the overlap line flux or continuum flux density ratios were forced to unity.
GOODSN-17940 was not an outlier, and was thus scaled by the median value appropriate for its mask.
After applying the grating-to-grating scaling, a comparison of the fluxes of 156 emission lines across the full AURORA sample
 with S/N$\ge$5 measured in the overlap
 region between two gratings yields a median offset of 0.1\%\ with an intrinsic scatter of 8\%,
 demonstrating that the systematic uncertainty on flux calibration between gratings is $<10\%$ for AURORA.

After correcting the relative flux calibration between gratings, the final absolute flux calibration is determined using the following method.
For each target, the 1D spectrum in each grating was passed through the transmission curves
 of all available {\it HST} and {\it JWST}/NIRCam imaging filters fully overlapping in wavelength coverage with that grating to produce mock photometry.
The flux density ratios of the imaging photometry to mock photometry were calculated for each filter in which both the mock and imaging
 photometry have S/N$\ge$3.
The 1D spectrum in all 3 gratings was then multiplied a single scale factor defined by the median of these ratios,
 effectively tying the final absolute flux calibration of the 1D spectra to the imaging photometry.
A different absolute scale factor was thus calculated for each AURORA target using its 1D spectra and imaging photometry.
Across all AURORA targets, the median absolute scale factor was 1.35, with a standard deviation of 0.17~dex.

\subsection{Photometric Catalogs and SED Fitting}\label{sec:sed}

We make use of the photometric catalogs from the DAWN {\it JWST} Archive\footnote{\url{https://dawn-cph.github.io/dja/}} that provides measurements in the available {\it HST}/WFC3, {\it HST}/ACS, and
 {\it JWST}/NIRCam imaging filters sampling the spectral energy distributions (SEDs) of each target \citep{val2023}.
In the GOODS-N field, this catalog includes: {\it JWST}/NIRCam imaging in the F090W, F115W, F150W, F182M, F200W, F210M, F277W, F335M, F356W, F410M, and F444W
 filters from JADES \citep{eis2023};
 {\it HST}/WFC3 imaging in the F606W, F775W, F814W, F850LP, F105W, F125W, F140W, and F160W filters;
 and {\it HST}/ACS imaging in the F435W filter. 
The combined filter set spans a rest-frame wavelength range of $700-9000$~\AA\ for GOODSN-17940.

\begin{figure}
\centering
\includegraphics[width=1.0\columnwidth]{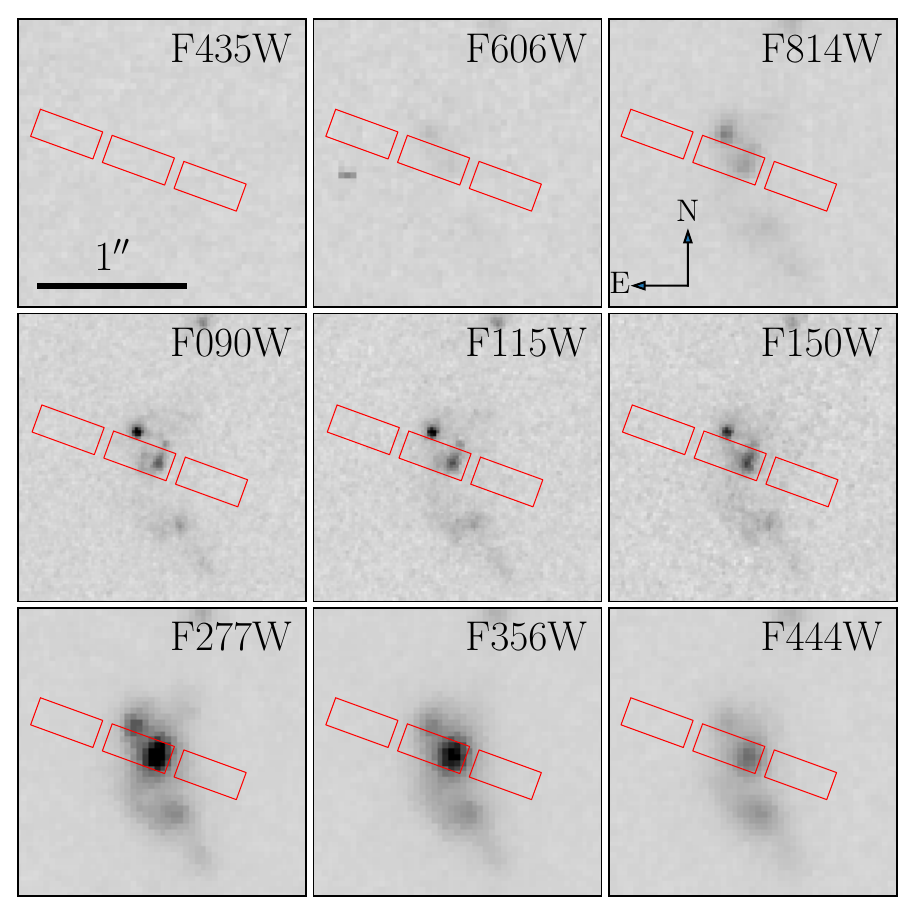}
\caption{
{\it JWST}/NIRCam 
broadband imaging of GOODSN-17940 in F453W, F606W, and F814W from {\it HST} ACS; and F090W, F115W, F150W, F277W, F356W, and F444W from {\it JWST}/NIRCam.
Images are in order of increasing wavelength from top left to bottom right, spanning $\lambda_{\mathrm{rest}}\approx 700-9,000$~\AA.
Each cutout is centered on the same position, $2^{\prime\prime}$ in side length, oriented with north up and east left, and
 displayed with a linear stretch where $F_\nu$ per arcsec$^2$ is the same in all panels.
The red rectangles show the locations of the NIRSpec MSA microshutter slitlet in the central position of the 3-point nod pattern.
}\label{fig:nircam}
\end{figure}

Figure~\ref{fig:nircam} shows broadband imaging of GOODSN-17940 in a selection of 3 {\it HST} and 6 {\it JWST}/NIRCam filters,
 with the position of the NIRSPec microshutter slitlet overlaid.
GOODSN-17940 appears to be part of a potentially merging system, where the AURORA slitlet predominantly captures light from the central of
 three components.
The extended component offset $\approx$$0.5^{\prime\prime}$ to the SW was observed with NIRSpec in the prism, $R\sim1000$, and $R\sim2700$ modes by JADES \citep[PID: 1181][]{eis2023,deu2025}
 and was confirmed to be at $z_\mathrm{spec}=4.4103$.
This SW component is thus physically close to GOODSN-17940 and likely gravitationally interacting, as suggested by the tidal-tail like features.
The compact component offset $\approx$$0.3^{\prime\prime}$ to the NE does not have a published spectroscopic redshift, but appears to be at a similar redshift
 to the two other components since all three display a photometric dropout between F606W and F435W (the Lyman break is at $\approx$5,000~\AA\ at $z=4.41$).
GOODSN-17940 is thus possibly the central component of a triple galaxy merger, which may have triggered the strong starburst that this source is undergoing
 based on the large equivalent widths of its nebular emission lines.
The strong [O\iii]$\lambda\lambda$4960,5008+H$\beta$ and H$\alpha$ lines are responsible for the excess brightness in the F277W and F356W filters relative
 to the other filters.

The models used for SED fitting only incorporate stellar continuum emission.
Accordingly, the measured photometric flux densities require corrections for contributions from nebular line and continuum emission before fitting.
Nebular continuum emission was modeled using Cloudy \citep{fer2017} with an input ionizing spectrum from a 100~Myr old
 BPASS binary stellar population \citep{eld2017} under the assumption of a constant star-formation history.
The gas density was set to $250~\mathrm{cm}^{-3}$ appropriate for the typical densities measured at $z\sim2-3$ \citep[e.g.,][]{san2016,str2017}.
Since the nebular continuum shape is not sensitive to the ionization parameter $U$ \citep{byl2017}, we adopt a fixed log($U)=-2.5$.
The nebular metallicity $Z_\mathrm{neb}$, primarily tracing the gas-phase O/H, was varied from $0.05-1.5~Z_\odot$ (12+log(O/H$)=7.4-8.9$).
The stellar metallicity $Z_*$ was treated as a proxy of the Fe abundance to which the shape of the ionizing spectrum is sensitive.
At each value of nebular metallicity, we adopted $Z_*=Z_\mathrm{neb}/4$ equivalent to assuming 4 times solar $\alpha$-enhancement
 based on observational constraints on the O/Fe abundance ratios of $z\sim2-3$ star-forming galaxies \citep[e.g.,][]{ste2016,str2018,sha2019,top2020a,top2020b,san2020,cul2021,sta2024}.
For each of the resulting Cloudy models, we extracted the pure nebular continuum emission normalized to the H$\beta$ line intensity.

The nebular corrections to the photometry were calculated using the following method. 
First, a model nebular emission spectrum was produced over observed wavelengths of $9000~\mathrm{\AA}-5.2~\mu\mathrm{m}$.
The model spectrum was populated with emission lines by adding Gaussian profiles using the best-fit centroids, fluxes, and widths
 from the emission line fitting (Sec.~\ref{sec:emline}) for all lines detected at $S/N\ge5$.
An initial estimate of the gas-phase metallicity was then obtained using strong-line ratios of [O\ii]$\lambda\lambda$3727,3730,
 [Ne\iii]$\lambda$3870, H$\beta$, and [O\iii]$\lambda$5008 in combination with the \citet{bia2018} high-redshift analog calibration set.
The Cloudy model most closely matched in $Z_\mathrm{neb}$ to this metallicity was chosen to obtain the nebular continuum model appropriate for each galaxy.
The H$\alpha$/H$\beta$ ratio was used to infer the nebular reddening, \ebvgas, assuming a \citet{car1989} Milky Way extinction curve,
 and the H$\beta$ flux was dust corrected.
The Cloudy nebular continuum model (normalized to the H$\beta$ intensity) was then multiplied by the dust-corrected H$\beta$ flux to
 obtain the intrinsic nebular continuum in flux density units.
Finally, the intrinsic nebular continuum was reddened according to the derived \ebvgas\ and the \citet{car1989} curve to obtain a model
 of the observed nebular continuum flux density.
This final nebular continuum model was then added to the nebular emission model spectrum.
For each photometric filter, the model spectrum was passed through the corresponding transmission curve and the resulting nebular model flux density was
 subtracted from the measured flux density to obtain the corrected flux density.
Uncertainties on the nebular correction were obtained by perturbing the measured emission-line fluxes according to their errors and repeating the
 above process 500 times, adopting half of the 16th-84th percentile of the resulting distribution as the $1\sigma$ error.
This nebular component error was added in quadrature to the error on the photometry.

GOODSN-17940 has very strong emission lines such that the nebular correction is crucial to obtaining
 an accurate stellar population model.
The rest-frame equiavelent widths of H$\alpha$, H$\beta$, and [O\iii]$\lambda$5008 measured from the AURORA spectra
 are $1630\pm40$~\AA, $280\pm10$~\AA, and $1370\pm20$~\AA, respectively.
Consequently, the emission-line corrections are significant,
 changing the photometry in filters covering H$\alpha$ and [O\iii]$\lambda$5008 by $\approx1.5$~mag.
Furthermore, the nebular continuum contribution is important, making up $10-50\%$ of the continuum flux density, with
 the strongest contributions just bluewards of the Balmer ($\lambda_\mathrm{rest}=3646$~\AA) and Paschen ($\lambda_\mathrm{rest}=8201$~\AA) limits.

Stellar population parameters and a model of the stellar continuum were obtained by fitting
 the corrected photometry with flexible stellar population synthesis (FSPS) models \citep{con2009} using FAST \citep{kri2009},
 assuming a \citet{cha2003} initial mass function (IMF) and delayed-$\tau$ star-formation histories ($SFR(t)\propto t e^{-t/\tau}$),
 where the age $t$ is the time since the onset of star formation.
Following \citet{red2018}, each AURORA target was fit under two assumptions about the metallicity and dust attenuation curve:
 $Z_*=0.019$ ($1.4~Z_\odot$) with a \citet{cal2000} curve (the ``Calzetti'' model)
 and $Z_*=0.0031$ ($0.2~Z_\odot$) with a \citet{gor2003} SMC curve (the ``SMC'' model).
The choice that yielded a lower $\chi^2$ value for each target was adopted to obtain fiducial stellar population parameters and the stellar
 continuum model used in emission-line fitting (Sec.~\ref{sec:emline}).
For GOODSN-17940, the SMC model yields a stellar mass of log($M_*/\mathrm{M}_\odot)=9.01\pm0.03$, SFR(SED$)=38\pm2\ \mathrm{M}_\odot~\mathrm{yr}^{-1}$, $t_\mathrm{age}=50\pm9$~Myr, $\tau=100$~Myr,
 and $E(B-V)_{\text{stars}}=0.11\pm0.01$.
With the Calzetti model, we obtain log($M_*/\mathrm{M}_\odot)=8.92\pm0.05$, SFR(SED$)=132\pm50\ \mathrm{M}_\odot~\mathrm{yr}^{-1}$, $t_\mathrm{age}=13\pm6$~Myr, $\tau=100$~Myr,
 and $E(B-V)_{\text{stars}}=0.26\pm0.04$.
While both models yield a similar stellar mass and a rising star-formation history($t_\mathrm{age}<\tau$), the Calzetti model has a larger SFR, a younger age, and more reddening.
For GOODSN-17940, the $Z_*=0.0031$ and SMC curve case yielded a lower $\chi^2$ and the stellar continuum from this model was used in emission line fitting.

To check whether the choice of the delayed-$\tau$ parameterization impacted the derivation of the stellar population properties, we also
 fit the photometric SED of GOODSN-17940 using a non-parametric star-formation history with PROSPECTOR \citep{jon2021}
 using eight indendent time bins, six of which are evenly logarithmically spaced from the Big Bang to 10~Myr lookback time
 and the final two of which span $0-3$~Myr and $3-10$~Myr lookback times.
This setup allows the SFH to capture a recent burst, as indicated by the large H$\alpha$ and H$\beta$ equivalent widths,
 independently from any star formation at earlier times in a way that the delayed-$\tau$ form cannot.
However, we find that the non-parametric PROSPECTOR results are in good agreement with those described above,
 yielding log($M_*/\mathrm{M}_\odot)=8.79\pm0.01$, SFR(SED$)=87\pm5\ \mathrm{M}_\odot~\mathrm{yr}^{-1}$ over the past 10~Myr, mass-weighted age $t_\mathrm{age}=7.1_{-0.6}^{+14.7}$~Myr, 
 and $E(B-V)_{\text{stars}}=0.29\pm0.01$.
Results from the delayed-$\tau$ and non-parametric SFH models thus both  indicate GOODSN-17940 is dominated by an extremely young stellar population. 

\subsection{Emission Line Measurements}\label{sec:emline}

\begin{figure*}
\centering
\includegraphics[width=\textwidth]{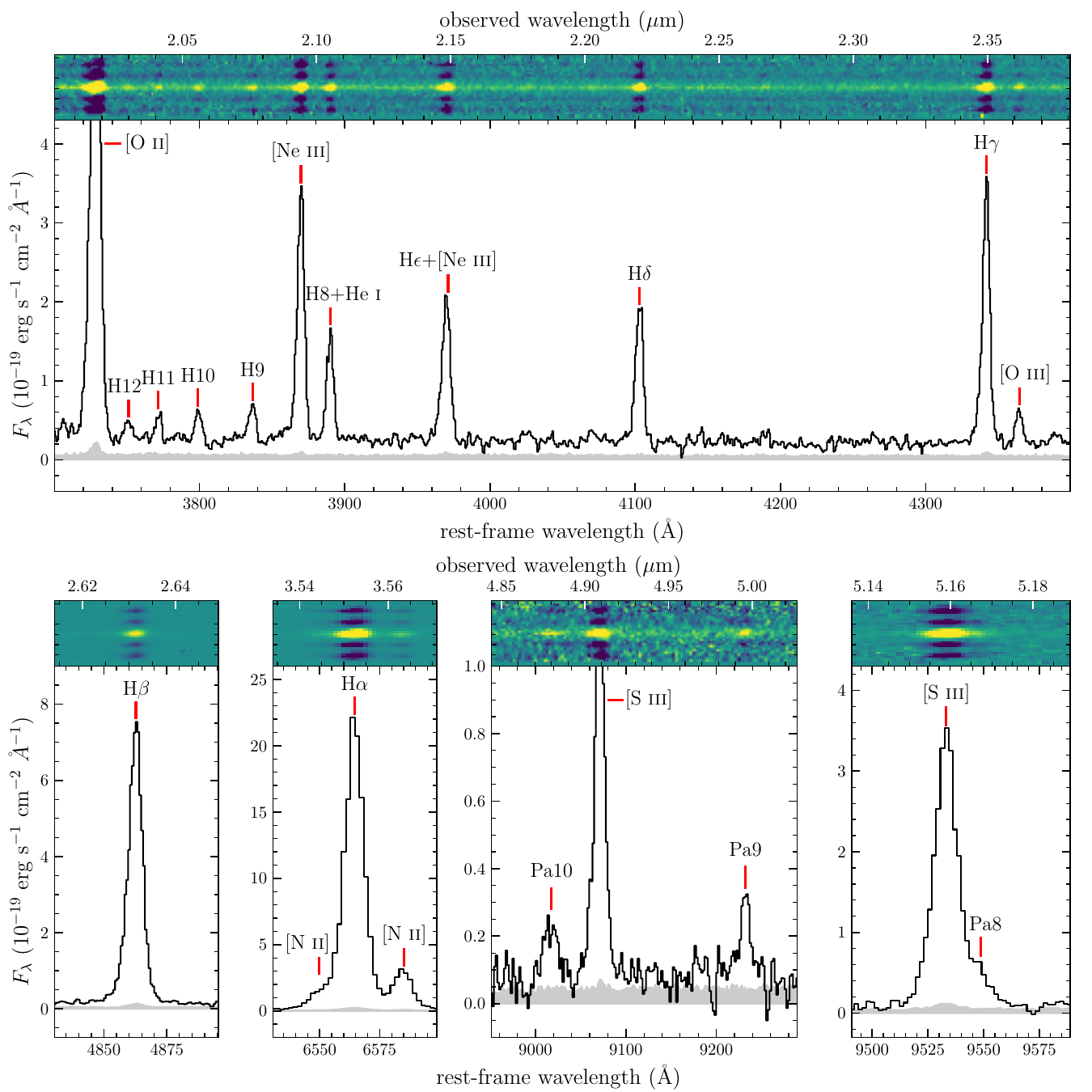}
\caption{Detected H\one\ emission lines in the spectrum of GOODSN-17940.
In each panel, the 2D spectrum is shown at the top and the 1D science spectrum is displayed at the bottom.
Emission lines detected with $S/N\ge5$ are labeled.
Two detected H\one\ lines, H8 and H$\epsilon$, are blended with nearby emission lines due to the instrumental resolution.
Pa8 is partially blended with [S\iii]$\lambda$9533 due to the intrinsic line widths of GOODSN-17940, but the two
 lines have distinct peaks and are cleanly separable with a double Gaussian fit.
The measured line fluxes of the 13 detected H\one\ lines are reported in Table~\ref{tab:fluxes}.
Lines spanning H12 to H$\beta$ fell in the G235M grating, while H$\alpha$ and the Paschen-series lines were observed with the G395M grating.
}\label{fig:spec}
\end{figure*}

Measurements of emission line fluxes and redshifts were obtained by fitting Gaussian profiles to the 1D science spectra.
Initial redshifts were inferred by fitting Gaussian profiles to the brightest emission line in each grating during 1D extraction.
The final redshift was measured by fitting a single Gaussian profile to the highest S/N line among all 3 gratings,
 which was typically [O\iii]$\lambda$5008 or H$\alpha$ (the case for 79 out of 95 AURORA targets with robust redshifts).
When fitting the brightest line for the redshift, we also inferred the instrument-corrected velocity width of this line.

The spectral resolution for each grating reported in the Jdox\footnote{Available at
 \url{https://jwst-docs.stsci.edu/jwst-near-infrared-spectrograph/nirspec-instrumentation/nirspec-dispersers-and-filters}.}
 is appropriate for a fully and evenly illuminated microshutter.
This scenario is not appropriate for our targets, the majority of which are spatially extended but do not typically come close to filling the microshutter.
Consequently, the appropriate spectral resolution for our targets will be higher than the $R\sim1000$ reported in JDox for the medium NIRSpec gratings.
For each AURORA target, new effective spectral resolutions in G140M, G235M, and G395M were calculated using \texttt{msafit} \citep{deg2024}.
This software package can take an intrinsic morphology model as input to forward model the effective NIRSpec spectral resolution as a function of wavelength.
Morphological models in each grating were obtained by fitting one or more PSF-convolved 2D S\'{e}rsic profiles to imaging in a NIRCam filter
 falling within the NIRSpec grating wavelength coverage (F115W for G140M, F277W for G235M, F444W for G395M).
The new spectral resolutions were typically $1.3-1.6$ times higher than those reported in JDox for a uniformly illuminated microshutter.
For GOODN-17940, the new spectral resolutions are $R\sim1400$ at the central wavelengths of the G140M, G235M, and G395M bandpasses.

Emission lines were fit with Gaussian profiles, for which the instrument-corrected velocity width was constrained to be
 within 20\%\ of the value measured for the bright line from which the redshift was measured.
Closely spaced lines with $\Delta\lambda/\lambda<0.01$ (e.g., H$\alpha$ and [N\ii]$\lambda\lambda$6550,6585) were fit
 simultaneously with multiple Gaussians, though some closely-spaced lines (e.g., the high-order Balmer lines
 near [O\ii]$\lambda$3728 and [Ne\iii]$\lambda$3870) were manually split up to prevent large numbers ($N>4$) of lines
 from being fit simultaneously.
Closely-spaced doublets from the same ion (e.g., [S\ii]$\lambda\lambda$6718,6733; [N\ii]$\lambda\lambda$6550,6585) were constrained to have the same line width.

The continuum under the lines was taken to be the sum of the best-fit stellar population model from SED fitting
 and the reddened nebular continuum model computed during photometric correction (Sec.~\ref{sec:sed}).
This continuum model was convolved with a wavelength-dependent Gaussian kernel to match the instrumental resolution at the
 wavelength of each fitted line.
When fitting each line or set of lines, the continuum model was allowed to vary by a multiplicative factor to
 fine tune the continuum level locally, though we note that these corrections are small as the flux calibration
 of the spectrum is tied to the photometry (Sec.~\ref{sec:fluxcal}).
Including the stellar population model during fitting naturally corrects the measured line fluxes for any
 underlying absorption features such as those affecting H\one\ Balmer lines.
Since the SED fitting depends on corrections to the photometry based on measured line fluxes,
 the emission line measurements and SED fitting were performed iteratively.
A first pass of emission line fitting was carried on using the stellar population model based on the
 raw uncorrected photometry.
The resulting line flux measurements were then used to perform the nebular emission correction to the photometry,
 the SED fitting was re-run,
 and this second-pass best-fit stellar population was adopted for the continuum model to measure the final
 emission line parameters.
Uncertainties on the best-fit emission-line parameters were estimated by perturbing the 1D spectra according to the error spectra,
 re-measuring all emission lines, repeating this process 500 times to build up a distribution,
 and taking half of the 16th-84th percentile width to be the $1\sigma$ error on each property.

This work focuses on the dust properties of GOODSN-17940 derived from 13 H\one\ recombination lines
 detected at $S/N\ge5$ (10 Balmer and 3 Paschen),
 the measured fluxes of which are presented in Table~\ref{tab:fluxes} 
 alongside the line flux uncertainty estimates described above.
The 1D and 2D spectra covering the wavelengths of these 13 H\one\ lines are shown in Figure~\ref{fig:spec}.
We detect high-order Balmer lines up to H12 at $\lambda_\mathrm{rest}=3751$~\AA\ ($S/N=6.7$).
It was not possible to measure further Balmer series lines due to blending with the much brighter [O\ii]$\lambda\lambda$3727,3730 doublet.
The wide wavelength coverage of AURORA enables simultaneous coverage of three Paschen series lines in the near-infrared:
 Pa10 to Pa8 at $\lambda_\mathrm{rest}=9018-9550$~\AA.

\begin{table}
 \centering
 \caption{Measured H\one\ line fluxes for GOODSN-17940.
 }\label{tab:fluxes}
 \begin{tabular}{ l l l }
\hline\hline
Line  &  $\lambda_\mathrm{rest}$  &  F$_{obs}(\lambda$)  \\ 
 & {\scriptsize \AA\ (vac.)} & {\scriptsize $10^{-18}~\mbox{erg s}^{-1}~\mbox{cm}^{-2}$} \\
\hline
H12  &  3751.22  &  $0.74\pm0.11$  \\
H11  &  3771.70  &  $0.79\pm0.10$  \\
H10  &  3798.98  &  $1.09\pm0.10$  \\
H9  &  3836.48  &  $1.56\pm0.11$  \\
H8+He\one\ $\lambda$3890\tablenotemark{a}  &  3890.17  &  $4.25\pm0.11$  \\
H$\epsilon$+$[$Ne\iii$]$$\lambda$3969\tablenotemark{a}  &  3971.20  &  $7.13\pm0.14$  \\
H$\epsilon$\tablenotemark{b} &  3971.20  & $3.82\pm0.31$  \\
H$\delta$  &  4102.89  &  $6.36\pm0.11$  \\
H$\gamma$  &  4341.68  &  $10.95\pm0.12$  \\
H$\beta$  &  4862.68  &  $26.90\pm0.17$  \\
H$\alpha$  &  6564.61  &  $115.70\pm0.55$  \\
Pa10  &  9017.77  &  $1.32\pm0.16$  \\
Pa9  &  9232.23  &  $1.86\pm0.21$  \\
Pa8  &  9548.82  &  $2.51\pm0.29$  \\
\hline
\end{tabular}
\tablenotetext{a}{The sum of the flux is reported for these blended lines.}
\tablenotetext{b}{Deblended from [Ne\iii]$\lambda$3969 using the measured [Ne\iii]$\lambda$3870 flux.}
\end{table}

\section{Results} \label{sec:results}

\subsection{The Optical-to-near-infrared Nebular Attenuation Curve}\label{sec:opt}

To derive the nebular attenuation curve spanning optical to near-infrared wavelengths,
 we utilize the H\one\ emission lines that are detected with a signal-to-noise ratio S/N$\ge$5,
 satisfied by 13 emission lines: H$\alpha$ through H12 (3751~\AA) from the Balmer series
 and Pa8 (9550~\AA) to Pa10 (9018~\AA) from the Paschen series (Fig.~\ref{fig:spec}).
Of these 13 lines, two are blended with neighboring lines at the spectral resolution of the medium NIRSpec gratings.
H$\epsilon$ is blended with [Ne\iii]$\lambda$3969 and H8 is blended with He\one$\lambda$3890.
We deblended H$\epsilon$ by subtracting the predicted [Ne\iii]$\lambda$3969 flux from the total flux measured across the blended feature
 using the measured [Ne\iii]$\lambda$3870 flux and the fixed intrinsic ratio [Ne\iii]$\lambda$3870/$\lambda$3969=3.32 calculated
 with Pyneb \citep{lur2015}.
We exclude H8 from the remaining analysis to avoid uncertainties related to predicting the He\one\ spectrum.
The remaining H\one\ recombination lines are isolated or cleanly separated from nearby lines, including Pa8
 that falls near [S\iii]$\lambda$9533 but is separated by $>$2 resolution elements from the [S\iii] centroid and
 is robustly deblended by fitting the two lines simultaneously.
We thus include 12 H\one\ lines in this analysis spanning rest-frame wavelengths of 3751~\AA\ to 9550~\AA.

We follow the procedure described in \citet{red2020} to calculate the shape of the nebular attenuation curve.
The intrinsic flux (\fint) and observed attenuated flux (\fobs) of a line at rest-frame wavelength $\lambda$ are related according
 to the magnitudes of attenuation at that wavelength ($A(\lambda)$) by
\begin{equation}
f_\mathrm{int}(\lambda) = f_\mathrm{obs}(\lambda) \times 10^{0.4 A(\lambda)}
\end{equation}
The attenuation at two wavelengths $\lambda_1$ and $\lambda_2$ can be related according to
\begin{equation}
A(\lambda_2) = 2.5\left[ \mathrm{log}_{10}\left(\frac{f_\mathrm{obs}(\lambda_1)}{f_\mathrm{obs}(\lambda_2)}\right) - \mathrm{log}_{10}\left(\frac{f_\mathrm{int}(\lambda_1)}{f_\mathrm{int}(\lambda_2)}\right) \right] + A(\lambda_1)
\end{equation}
We normalize this equation such that $A(\lambda_1)=1$, obtaining
\begin{equation}\label{eq:aprime}
A^\prime(\lambda_2) = 2.5\left[ \mathrm{log}_{10}\left(\frac{f_\mathrm{obs}(\lambda_1)}{f_\mathrm{obs}(\lambda_2)}\right) - \mathrm{log}_{10}\left(\frac{f_\mathrm{int}(\lambda_1)}{f_\mathrm{int}(\lambda_2)}\right) \right] + 1
\end{equation}
where $A^\prime(\lambda_2)=A(\lambda_2) + [1 - A(\lambda_1)]$.
We take the reference $\lambda_1$ to be the wavelength of the H$\beta$ line at $\lambda_\mathrm{rest}=4863$~\AA,
 and express $\lambda_2$ as $\lambda$ hereafter.

\begin{figure*}
\centering
\includegraphics[width=0.495\textwidth]{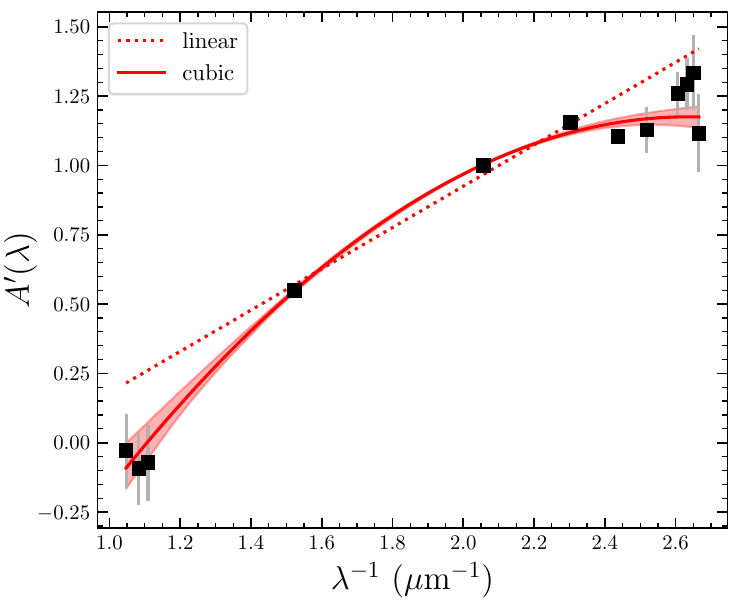}
\includegraphics[width=0.495\textwidth]{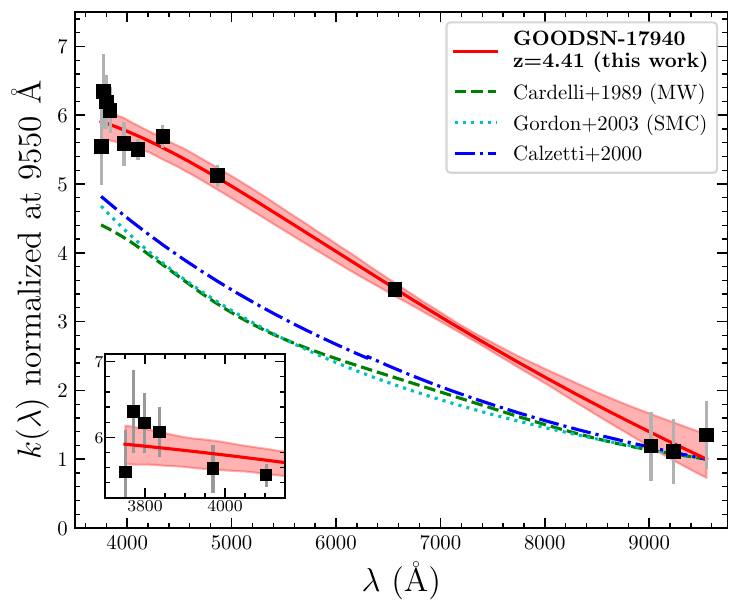}
\caption{\textbf{Left:} $A^\prime(\lambda)$ vs. $1/\lambda$.
The black squares show the values of $A^\prime$ calculated from the measured H\one\ flux ratios relative to H$\beta$ using equation~\ref{eq:aprime}.
The dotted red line displays a fit assuming a functional form that is linear in $\lambda^{-1}$, which fails to match the data.
The solid red line shows the best-fit cubic-in-$\lambda^{-1}$ function that we use for this analysis.
The light red shaded region shows the $1\sigma$ bounds on the best-fit cubic.
\textbf{Right:} Best-fit nebular attenuation curve $k(\lambda)$ of GOODSN-17940 at optical and near-infrared wavelengths is
 shown by the solid red line, derived using equation~\ref{eq:kprime}.
The Milky Way extinction curve \citep{car1989}, SMC extinction curve \citep{gor2003}, and $z\sim0$ starburst attenuation curve \citep{cal2000},
 are shown for comparison.
All curves have been normalized such that the value at 9550~\AA\ is unity.
}\label{fig:optcurve}
\end{figure*}

We calculate $A^\prime(\lambda)$ for the 12 H\one\ lines in our sample and show the results as a function of $\lambda^{-1}$
 in the left panel of Figure~\ref{fig:optcurve}.
The intrinsic brightness ratios were calculated using the Python package Pyneb \citep{lur2015} assuming an electron
 temperature $T_e=11,000$~K and density $n_e=900~\mathrm{cm}^{-3}$, the approximate values inferred from the auroral
 [O\iii]$\lambda$4364 line and the [S\ii]$\lambda\lambda$6718,6733 doublet (Sanders et al., in prep.).
\citet{red2020} fit their $A^\prime(\lambda)$ values with linear and quadratic functional forms in $\lambda^{-1}$,
 however we find that these forms fail to capture the curvature displayed by the $A^\prime(\lambda)$ values for GOODSN-17940.
We found that a cubic function in $\lambda^{-1}$ was the lowest-order polynomial that provided a good fit, shown by the
 solid red line in Fig.~\ref{fig:optcurve}, left.
It is notable that the \citet{car1989} Milky Way extinction curve, \citet{cal2000} starburst attenuation curve, and the
 \citet{gor2003} SMC extinction curve are all well-approximated
 by a form that is linear in $\lambda^{-1}$ at optical wavelengths, indicating that the nebular attenuation curve of GOODSN-17940 deviates
 significantly from these commonly assumed curves in high-redshift galaxy studies.
Even if we limit the lines considered to the Balmer series, a linear functional form still cannot simultaneously fit both H$\alpha$ and
 the high order Balmer lines.
Accordingly, the deviation from a Milky Way-like curve does not solely depend on the inclusion of the rest-frame near-infrared Paschen series lines.

The value of the attenuation curve $k(\lambda)$ is defined using the ``total'' definition in which the absolute $A(\lambda)$ is normalized by the color
 excess between the $B$ and $V$ bands at 4400~\AA\ and 5500~\AA, respectively:
\begin{equation}
k(\lambda) = \frac{A(\lambda)}{E(B-V)} = \frac{A(\lambda)}{A(4400\mathrm{~\AA}) - A(5500\mathrm{~\AA})}
\end{equation}
For the nebular attenuation curve, the color excess is the reddening for the ionized gas component, \ebvgas.
The attenuation curve is related to the normalized $A^\prime(\lambda)$ by a normalization constant $C$ that is independent of wavelength:
\begin{equation}\label{eq:kprime}
k(\lambda) - C = k^\prime(\lambda) = \frac{A^\prime(\lambda)}{A^\prime(4400\mathrm{~\AA}) - A^\prime(5500\mathrm{~\AA})}
\end{equation}
with $C = [A(\lambda_1) - 1]/E(B-V)_{\text{gas}}$.
The wavelength-dependent shapes of $k(\lambda)$ and $k^\prime(\lambda)$ are thus identical.
Adopting the convention that the value of the attenuation curve $k(\lambda)$ at 5500~\AA\ is $R_V=A(V)/E(B-V)$,
 we can derive the value of the normalization constant as a function of $R_V$: $C=R_V-k^\prime(5500~\mathrm{\AA})$.

The resulting nebular attenuation curve $k(\lambda)$ is shown in the right panel of Figure~\ref{fig:optcurve} normalized to unity at the wavelength
 of the reddest H\one\ line available ($k^\prime(9550\mathrm{~\AA})=1$), derived from the best-fit cubic-in-$\lambda^{-1}$ form of $A^\prime(\lambda)$.
This curve was obtained by calculating $k^\prime(\lambda)$ for each available H\one\ line (black squares) and fitting these data points with
 a cubic-in-$\lambda^{-1}$ functional form, and converting to $k(\lambda)$ using $C=R_V-k^\prime(5500~\mathrm{\AA})$ yielding:
\begin{equation}\label{eq:kopt}
k_\mathrm{opt}(\lambda) = -13.730 + 13.572x - 4.063x^2 + 0.414x^3 + R_V
\end{equation}
with $x=1~\mu\mathrm{m}/\lambda$.
This curve is valid over virtually the entire rest-optical range and into the near-infrared, from 3750~\AA\ to 9550~\AA.
For comparison, we show the Milky Way, \citet{cal2000}, and SMC curves with the same normalization to unity at 9550~\AA.
The new curve is significantly steeper on average than all three of these existing curves over this wavelength range,
 particularly displaying a steeper slope at red wavelengths ($6000-9500$~\AA) while the slope is similar at blue wavelengths ($<5000$~\AA).
Using this new curve, we perform a simultaneous fit to the 12 H\one\ lines to derive the best-fit nebular reddening, $E(B-V)_\mathrm{gas}^\mathrm{New}=0.278\pm0.004$.

\subsection{The Ultraviolet Nebular Attenuation Curve}\label{sec:uv}

The AURORA NIRSpec observations provide wide and continuous wavelength coverage over $1-5.2~\mu$m in the observed frame,
 or rest-frame 1800~\AA\ to 1~$\mu$m at the redshift of GOODSN-17940.
Accordingly, the G140M spectrum covers a large portion of the rest-frame ultraviolet continuum from 1800~\AA\ to 3500~\AA.
The rest-UV continuum shape of star-forming galaxies is often described as a power law, $F_\lambda \propto \lambda^{\beta}$,
 typically measured over the approximate wavelength range $1300-2600$~\AA, excluding any filters or wavelength ranges in
 the spectrum that would cover Ly$\alpha$ \citep[e.g.,][]{cal1994,red2015,red2018,cal2021}.
The power-law slope, $\beta$, has been used as a measure of reddening under the assumption of an intrinsic blue slope in the range
 $\beta_\mathrm{int}\sim-2.0$ to $-2.8$, with values closer to $\lesssim-2.6$ expected for  
 very young and low-metallicity stellar populations \citep[e.g.,][]{red2015,shi2020,top2022,cul2024}.
Galaxies with more dust-reddening will have redder UV slopes (i.e., lower values of $\beta$), with the change in $\beta$ as a function
 of $E(B-V)$ related to the assumed attenuation curve.
Curves that are steeper in the UV, such as the SMC curve \citep{gor2003}, result in a larger change in $\beta$ from the intrinsic slope
 at fixed $E(B-V)$ relative to shallower UV curves like that of \citet{cal2000}.
The emitted UV continuum is a combination of starlight and nebular continuum emission, with the nebular component becoming increasingly
 important at young ages, making up to $\sim20-40\%$ of the total UV continuum at $\approx$2,000~\AA\ for a simple stellar population at $\lesssim5$~Myr \citep{rei2010,byl2017}.

If the age of the stellar population dominating the UV continuum emission
 is young enough ($\lesssim$10~Myr), then the starlight in this wavelength range
 is dominated by O and B stars that are spatially coincident with the H\ii\ regions.
Consequently, the UV continuum will be affected by approximately the same reddening experienced by the emission lines and
 the nebular attenuation curve will effectively determine
 the amount of reddening and attenuation
 the UV continuum receives, rather than a distinct stellar attenuation curve.
The star-formation history inferred from SED fitting (Sec.~\ref{sec:sed}) is rising under both dust curve assumptions (SMC and \citealt{cal2000}),
 generally consistent with a very young age.\footnote{The age $t_\mathrm{age}$ reported for the delayed-$\tau$ SFH models
 is the time since the onset of star formation.  Since $t_\mathrm{age}\ll\tau$ for both the SMC and Calzetti models considered here, the SFH is rising in both
 cases and the median stellar age is much smaller than the reported $t_\mathrm{age}$=50~Myr and 13~Myr, respectively.}
Further evidence comes from the emission line strengths.
GOODSN-17940 is an extreme emission-line galaxy (EELG) characterized by large emission-line equivalent widths that suggest a very young age.
The rest-frame equivalent widths ($\mathrm{EW}_0$) of H$\alpha$ and H$\beta$ measured from the spectrum (in which the rest-optical continuum is detected at
 median S/N$\approx$4 per pixel)
 are $1620\pm40$~\AA\ and $280\pm10$~\AA, respectively.

\begin{figure}
\centering
\includegraphics[width=0.95\columnwidth]{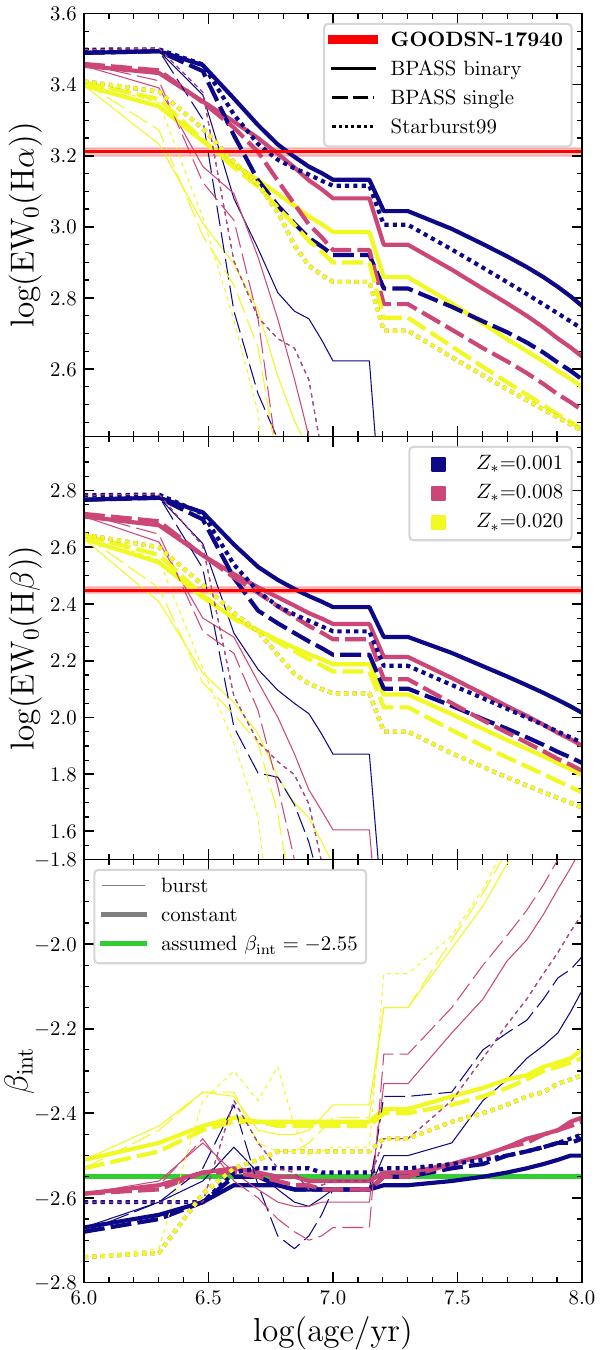}
\caption{Equivalent width of H$\alpha$ (top), equivalent width of H$\beta$ (middle), and intrinsic UV slope (bottom)
 as a function of age extracted from the Cloudy photoionization models described in Sec.~\ref{sec:uv}.
Thick lines represent models with a constant SFR, while thin lines indicate models with an instantaneous burst
 at $t=0$ and no further star formation.
Colors denote stellar metallicity.
The line style indicates which stellar models were used, including binary (solid) and single (dashed) BPASS models
 and Starburst99 models (dotted).
The red horizontal lines in the top two panels show the measured H$\alpha$ and H$\beta$ rest-frame equivalent widths for GOODSN-17940.
}\label{fig:cloudy}
\end{figure}

Figure~\ref{fig:cloudy} shows $\mathrm{EW}_0(\mathrm{H}\alpha)$, $\mathrm{EW}_0(\mathrm{H}\beta)$,
 and $\beta$ extracted from Cloudy photoionization models under a wide range of assumptions including:
 an incident spectrum from stellar populations of binary and single BPASS models \citep{eld2017} and Starburst99 models \citep{lei1999,lei2014};
 stellar metallicity ranging from $Z_*=0.001-0.020$ ($0.07-1.4~Z_\odot$) that sets the Fe abundance that most strongly
 impacts the UV continuum shape through absorption line blanketing;
 and a single instantaneous burst or a constant star-formation history.
We fix the ionization parameter at log($U)=-2.5$.
The exact assumed value of the ionization parameter $U$ has no effect on our results because
 the strength of the nebular continuum relative to the Balmer lines is insensitive to $U$ \citep{byl2017}.
We also assume no escape of H-ionizing Lyman continuum photons, i.e., $f_\mathrm{esc}=0$.
The nebular metallicity (i.e., gas-phase O/H) was set at 12+log(O/H$)=8.3$ ($0.4~Z_\odot$) to match
 the $T_e$-based direct metallicity derived from multiple auroral lines detected in the GOODSN-17940 spectrum (Sanders et al., in prep.).
The $\beta$ values were measured by fitting a power law to the combined stellar and nebular continuum over $1300-2600$~\AA\ after
 masking wavelength ranges where emission or absorption lines are present.
No dust was included in these Cloudy models, such that this slope is the intrinsic $\beta_\mathrm{int}$ with zero reddening.

In the top two panels of Figure~\ref{fig:cloudy}, it can be seen that the measured H$\alpha$ and H$\beta$ equivalent widths of GOODSN-17940
 can only be accommodated by models with very young ages between 2 and 6~Myr.
This conclusion is robust against the assumptions of the models since we have conservatively allowed a very wide range
 in stellar metallicity, star-formation history, and the input stellar models (BPASS vs.\ Starburst99).
Other independent sets of models have similarly found that such high equivalent widths require ages $<10$~Myr \citep[e.g.,][]{rei2010,xia2019}.
It is thus probable that GOODSN-17940 hosts a stellar population young enough that its rest-UV continuum emission will be
 attenuated according to the same curve that governs the nebular emission lines.
We can thus leverage a comparison of $\beta$ measured from the observed UV continuum to an assumed intrinsic UV slope to constrain
 the shape of the nebular attenuation curve at ultraviolet wavelengths.
At $<10$~Myr ages, the models have $\beta_\mathrm{int}=-2.7$ to $-2.4$, with an average value of $-2.55$ that we adopt as a fiducial value.

\begin{figure*}
\centering
\includegraphics[width=1.0\textwidth]{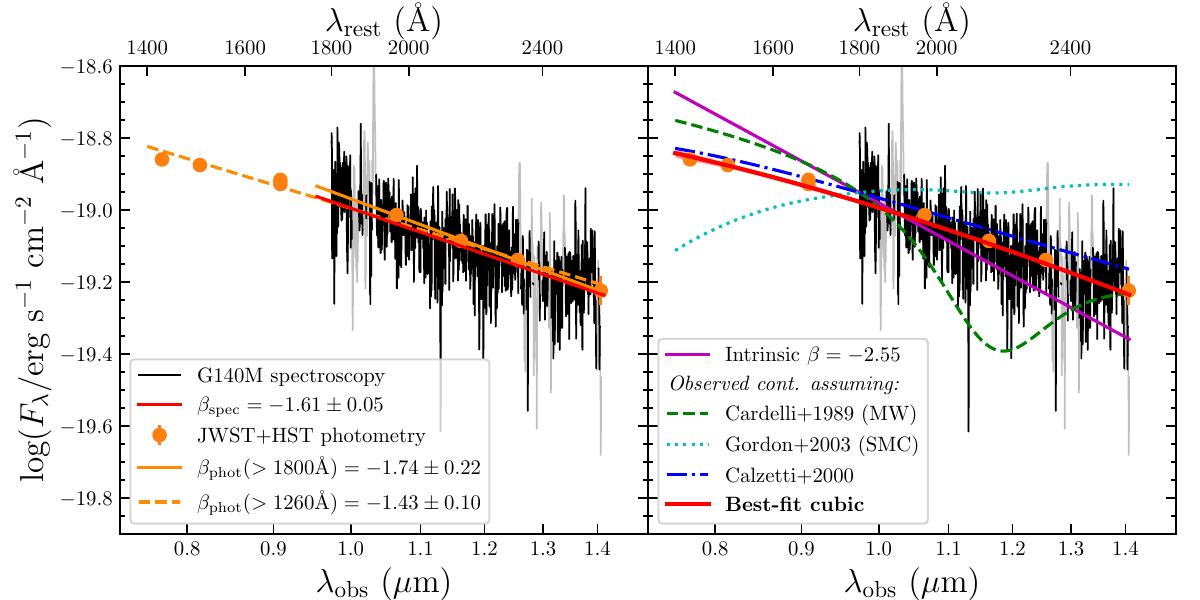}
\includegraphics[width=0.5\textwidth]{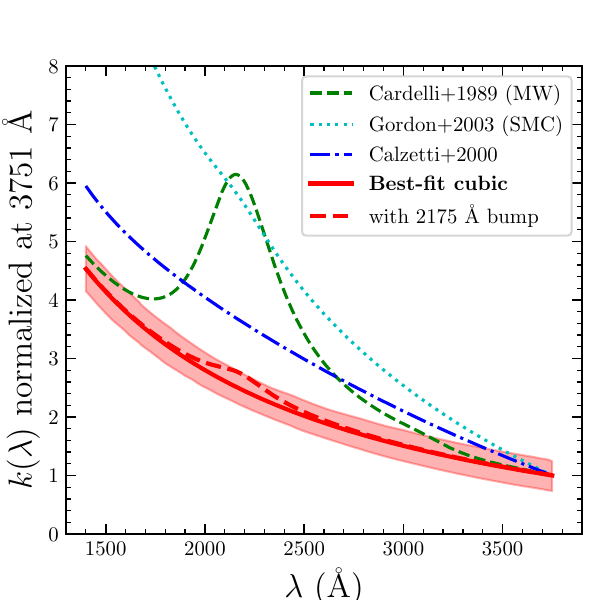}
\caption{\textbf{Top Left:} Rest-frame ultraviolet continuum of GOODSN-17940.
The NIRSpec G140M spectrum is shown in black, with masked emission and absorption lines displayed in gray.
The photometric measurements from {\it JWST} and {\it HST} imaging are presented as orange circles.
The best-fit power laws to the photometry (orange lines) and spectrum (red line) yield a UV slope of $\beta\sim-1.6$, suggesting a large amount of dust reddening.
\textbf{Top Right:} Inferred shape of the reddened UV continuum assuming an intrinsic spectrum that is a power law with an index of $\beta=-2.55$ (purple line).
The other lines show the shape of the UV continuum if this intrinsic continuum is reddened according to equation~\ref{eq:uvred} assuming a dust law that has a shape at UV wavelengths of
 the Milky Way extinction curve \citep[green dashed;][]{car1989}, the SMC extinction curve \citep[cyan dotted;][]{gor2003}, or the \citet{cal2000} attenuation curve
 (blue dash-dotted), none of which match the observed UV spectrum.
The red line shows the resulting reddened continuum shape using our best-fit cubic-in-$\lambda^{-1}$ UV attenuation curve (see bottom panel).
\textbf{Bottom:} Best-fit ultraviolet nebular attenuation curve of GOODSN-17940 (red solid line), alongside the Milky Way,
 SMC, and \citet{cal2000} curves for comparison.
All curves are normalized to unity at 3751~\AA.
The red dashed line shows the best-fit curve when a 2175~\AA\ bump component is added to the cubic functional form.
}\label{fig:uvcurve}
\end{figure*}

The top left panel of Figure~\ref{fig:uvcurve} shows the G140M spectrum from rest-frame 1800~\AA\ to 2600~\AA, as well as the photometry
 from {\it JWST} and {\it HST} imaging at $1260~\mathrm{\AA}<\lambda_\mathrm{rest}<2600~\mathrm{\AA}$.
The spectroscopy and photometry show remarkably good agreement in the overlapping wavelength range.
We fit a power law of the form $F_\lambda \propto \lambda^{\beta}$ to the G140M spectrum over $1800~\mathrm{\AA}<\lambda_\mathrm{rest}<2600~\mathrm{\AA}$,
 masking out ISM absorption and emission features, and obtain $\beta_\mathrm{spec}=-1.61\pm0.05$.
Fitting the rest-UV photometry results in $\beta_\mathrm{phot}(>1260~\mathrm{\AA})=-1.43\pm0.10$,
 or $\beta_\mathrm{phot}(>1800~\mathrm{\AA})=-1.74\pm0.22$ if we limit the photometry to the wavelength range covered by the G140M spectrum.
Both the spectroscopy and photometry present a consistent picture of a significantly reddened UV continuum relative to the expected
 $\beta_\mathrm{int}\approx-2.55$.
This result is in accord with the significant amount of dust reddening inferred from the H\one\ emission lines
 that yield $E(B-V)_\mathrm{gas}=0.278\pm0.004$ assuming our newly derived attenuation curve.

To determine the shape of the ultraviolet nebular attenuation curve, we assume an intrinsic UV continuum slope of $\beta_\mathrm{int}=-2.55$,
 the approximate average value from the Cloudy models described above,
 and infer the observed UV continuum by reddening the intrinsic continuum assuming a UV dust curve:
\begin{equation}\label{eq:uvred}
F_{\lambda,\mathrm{red}} \propto \lambda^{-2.55} \times 10^{-0.4 E(B-V)_\mathrm{gas} k_\mathrm{UV}(\lambda)}
\end{equation}
with \ebvgas\ set to the value inferred from the H\one\ lines.
The UV dust curve $k_\mathrm{UV}(\lambda)$ is normalized to match our new optical-to-NIR attenuation curve at 3751~\AA\ (equation~\ref{eq:kopt}).
We compare the resulting continuum shape to that measured from the spectroscopy and photometry of GOODSN-17940 to infer the shape of the nebular
 attenuation curve at UV wavelengths.

We begin by assuming that the dust curve has the same shape as the Milky Way, \citet{cal2000}, or SMC curve at $\lambda<3751$~\AA.
The resulting reddened UV continuum shapes are shown in the top right panel of Figure~\ref{fig:uvcurve}, normalized to match the flux density
 of the data at $\lambda_\mathrm{rest}=1800$~\AA.
It is clear that the SMC curve is much too steep to match the continuum shape of GOODSN-17940 since it produces a continuum that
 is far too red.
The Milky Way curve also yields a poor match to the data, in particular because of the strong bump present at 2175~\AA\ that results in
 excess attenuation over $1900-2500$~\AA.
The GOODSN-17940 data show no deviation from a flat power law over this wavelength range, suggesting that a 2175~\AA\ bump is not present
 in its attenuation curve.
The \citet{cal2000} starburst curve provides the closest match among these three, but still yields a redder continuum than is present in the data.

We derive the UV attenuation curve by fitting $F_{\lambda,\mathrm{red}}$ from equation~\ref{eq:uvred} to the combined
 spectroscopy and photometry of GOODSN-17940 between 1260~\AA\ and 2600~\AA\ after masking out emission and absorption features in the spectrum.
Due to the excellent agreement of the spectroscopy and photometry in the region of overlapping wavelength coverage, we do not apply
 any relative weighting between the two data types when fitting.
We assume a functional form that is cubic in $\lambda^{-1}$ for $k_\mathrm{UV}(\lambda)$,
 and impose the additional constaints that $k_\mathrm{UV}(3751~\mathrm{\AA})=k_\mathrm{opt}(3751~\mathrm{\AA})$ and the slope of
 $k_\mathrm{UV}$ matches that of $k_\mathrm{opt}$ at 3751~\AA\ such that the two will join smoothly.
The resulting ultraviolet attenuation curve is:
\begin{equation}\label{eq:kuv}
k_\mathrm{UV}(\lambda) = -0.535 + 0.706x + 0.009x^2 - 0.0001x^3 + R_V
\end{equation}
with $x=1~\mu\mathrm{m}/\lambda$.
Uncertainties on $k_\mathrm{UV}$ were inferred by perturbing $\beta_\mathrm{int}$ assuming a uniform distribution between $-2.7$ and $-2.4$,
 perturbing the data according to the measurement uncertainties, and repeating the fit 250 times.

The inferred reddened continuum assuming $\beta_\mathrm{int}=-2.55$ and our new $k_\mathrm{UV}$ is shown by the red line in the top right
 panel of Figure~\ref{fig:uvcurve}, and provides an excellent match to both the spectroscopy and photometry over $1400-2600$~\AA.
The bottom panel of Figure~\ref{fig:uvcurve} compares our new UV curve with the Milky Way, \citet{cal2000}, and SMC curves at $\lambda<3751$.
The new curve is shallower than both the SMC and \citet{cal2000} curves.
Interestingly, the new curve has a similar shape to that of the Milky Way curve in the absence of the 2175~\AA\ bump.
The dashed red line shows the result if we include a 2175~\AA\ bump component in our fit using a Drude profile centered at 2175~\AA\ with
 a width of 350~\AA\ \citep[e.g.,][]{fit1986,sal2018}, demonstrating that the best-fit bump strength is not significant relative to the uncertainties.
We thus significantly rule out the presence of a 2175~\AA\ bump in the GOODSN-17940 attenuation curve.

\subsection{The Combined Nebular Attenuation Curve from 1400~\AA\ to 9550~\AA}\label{sec:comb}

\begin{figure*}
\centering
\includegraphics[width=0.8\textwidth]{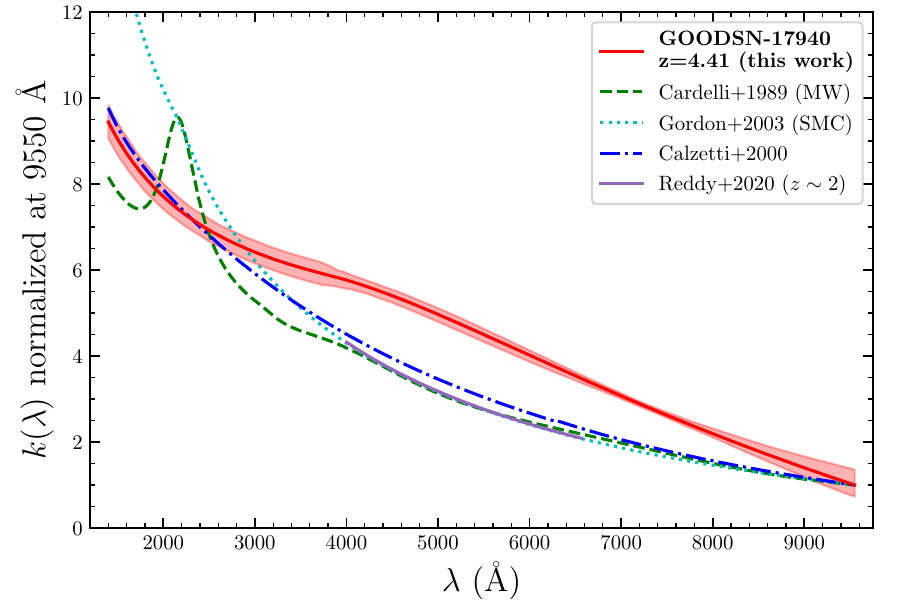}
\caption{The nebular attenuation curve of GOODSN-17940 from 1400~\AA\ to 9550~\AA\ (red line).
The Milky Way, SMC, \citet{cal2000}, and \citet{red2020} curves are shown for comparison.
All curves are normalized to unity at 9550~\AA.
}\label{fig:totcurve}
\end{figure*}

Here, we present the total functional form of the nebular attenuation curve of GOODSN-17940, a starburst galaxy at $z=4.41$, spanning
 from 1400~\AA\ to 9550~\AA:
\begin{equation}\label{eq:ktot}
\begin{aligned}
k(\lambda) &= -13.730 + 13.572x - 4.063x^2 + 0.414x^3 + R_V , \\ & \qquad\qquad\qquad 9550~\mathrm{\AA} \le \lambda \le 3751~\mathrm{\AA} ;
\\ &= -0.535 + 0.706x + 0.009x^2 - 0.0001x^3 + R_V , \\ & \qquad\qquad\qquad 3751~\mathrm{\AA} \le \lambda \le 1400~\mathrm{\AA} ;
\end{aligned}
\end{equation}
with $x=1~\mu\mathrm{m}/\lambda$.
The full curve is shown in Figure~\ref{fig:totcurve} in comparison to other commonly-assumed dust curves from the literature.
The uncertainty on the newly derived curve can be represented as an uncertainty in $k(\lambda)$ at fixed $\lambda$, with a median value of
 $\sigma_k=0.26$ at $\lambda<3751$~\AA\ and $\sigma_k=0.15$ at $\lambda>3751$~\AA.
The larger value of $\sigma_k$ in the UV is driven by uncertainty in our assumed intrinsic UV slope, $\beta_\mathrm{int}=-2.55\pm0.15$.
The nebular attenuation curve of GOODSN-17940 strongly deviates from the Milky Way extinction curve \citep{car1989},
 the SMC extinction curve \citep{gor2003}, and the starburst attenuation curve \citet{cal2000}.

The shape of this new curve is directly constrained between 3751~\AA\ and 9550~\AA\ by the observed flux ratios of 12 H\one\ emission lines,
 and between 1400~\AA\ and 2600~\AA\ based on the shape of the rest-UV continuum.
In contrast, the shape at $2600~\mathrm{\AA}<\lambda<3751~\mathrm{\AA}$ relies on extrapolating the best-fit UV curve out to 3751~\AA.
It is thus of interest to consider whether this extrapolation has properly set the relative normalization between 
 $k_\mathrm{opt}$ and $k_\mathrm{UV}$.
We can check this UV-to-optical normalization using the ratio of a UV line and an optical line that have a fixed intrinsic theoretical flux ratio,
 such that the difference between their intrinsic and observed flux ratio can be used to derive the difference in $k(\lambda)$ between the two wavelengths.

In the GOODSN-17940 spectrum, [O\ii]$\lambda$2471 and [O\ii]$\lambda\lambda$7322,7332 are both well detected, with respective significances of $9.9\sigma$ and $24.6\sigma$.
[O\ii]$\lambda$2471 is a blended doublet consisting of one line with an upper energy level of $n_\mathrm{up}=4$ and one with $n_\mathrm{up}=5$.
Similarly, [O\ii]$\lambda\lambda$7322,7332 is a quadruplet made up of two lines with $n_\mathrm{up}=4$ and two with $n_\mathrm{up}=5$.
Transitions originating from the same upper energy level have fixed brightness ratios.
The similar upper-level composition of these two [O\ii] multiplets thus leads to a fixed theoretical ratio of [O\ii]$\lambda$2471/[O\ii]$\lambda\lambda$7322,7332=0.75,
 calculated using the Python package Pyneb \citep{lur2015}.
The flux ratio measured from the spectrum is $0.30\pm0.03$.
If we dust correct this ratio adopting the total attenuation curve from equation~\ref{eq:ktot} and \ebvgas\ derived from the full suite of H\one\ lines,
 we obtain a dust-corrected ratio of $0.89\pm0.09$, consistent with the theoretical ratio at the $1.5\sigma$ level.
If the Milky Way curve of \citet{car1989} was instead used to derive \ebvgas\ and correct this [O\ii] ratio, the result is $1.89\pm0.21$,
 in $>5\sigma$ tension with the theoretical ratio.
This test implies that the extrapolation of the best-fit UV curve between 2600~\AA\ and 3751~\AA\ is reasonable, and the attenuation curve in equation~\ref{eq:ktot}
 is robust over the full $1400-9550$~\AA\ wavelength range.

We have left the normalization of the attenuation curve at 5500~\AA, $R_V$, as a free parameter.
Since our reddest available H\one\ line is at $<1~\mu$m, we cannot assume zero reddening at this wavelength.
Our adopted cubic-in-$\lambda^{-1}$ form also cannot be extrapolated to a longer wavelength where attenuation could be assumed to be
 zero because it is not monotonic nor does it asymptote to zero like a linear-in-$\lambda^{-1}$ form would (see \citealt{red2020}).
However, by requiring $k(9550~\mathrm{\AA})>0$ we obtain a lower limit of $R_V>3.50$.
A robust constraint on $R_V$ will require measurements of H\one\ lines at wavelengths long enough where attenuation can be assumed to be near zero,
 such as Pa$\alpha$ ($\lambda_\mathrm{rest}=1.876~\mu$m).
At $z=4.41$ for GOODSN-17940, Pa$\alpha$ falls at $10.1~\mu$m in the Channel 2 Long band for MIRI medium resolution spectroscopy.
It is thus possible to robustly establish the attenuation curve normalization for GOODSN-17940 with additional {\it JWST} observations.

\section{Discussion}\label{sec:discussion}

Using high signal-to-noise measurements of 12 H\one\ recombination lines and the rest-frame ultraviolet continuum,
 we have characterized the shape of the nebular dust attenuation curve from ultraviolet to near-infrared wavelengths
 ($1400~\mathrm{\AA}\le\lambda_\mathrm{rest}\le9550~\mathrm{\AA}$) for GOODSN-17940, a star-forming  galaxy at $z=4.41$.
This analysis was made possible by extremely deep (24~h on-source) spectroscopy that is continuous over $1-5~\mu$m
 (rest-frame $1800~\mathrm{\AA}-1~\mu$m) obtained as a part of the AURORA survey with {\it JWST}/NIRSpec.
The newly derived attenuation curve (Fig.~\ref{fig:totcurve}; eq.~\ref{eq:ktot}) significantly deviates from the most commonly assumed dust curves including
 the \citet{car1989} Milky Way extinction curve, the \citet{gor2003} SMC extinction curve, and the \citet{cal2000}
 attenuation curve of local starburst galaxies.
Specifically, it is steeper than all three of these curves at $\lambda\gtrsim5500$~\AA, similar in slope between 4000~\AA\ and 5500~\AA,
 but shallower than the SMC and \citet{cal2000} curves in the ultraviolet and lacks a 2175~\AA\ bump in contrast to the Milky Way curve.
Studies of the nebular attenuation curve of large samples of local galaxies spanning a wide range of stellar mass and SFR
 suggest that a Milky Way-like curve at optical wavelengths is relatively universal among star-forming galaxies \citep{rez2021,ji2023},
 such that the GOODSN-17940 curve generally differs from that of $z\sim0$ galaxies.
This deviation carries significant implications for inferences of the amount of dust reddening,
 star-formation rates from H\one\ emission lines,
 and physical properties from line ratios that require reddening corrections.

\subsection{Implications for reddening derived from different H\one\ lines}

\begin{figure}
\centering
\includegraphics[width=\columnwidth]{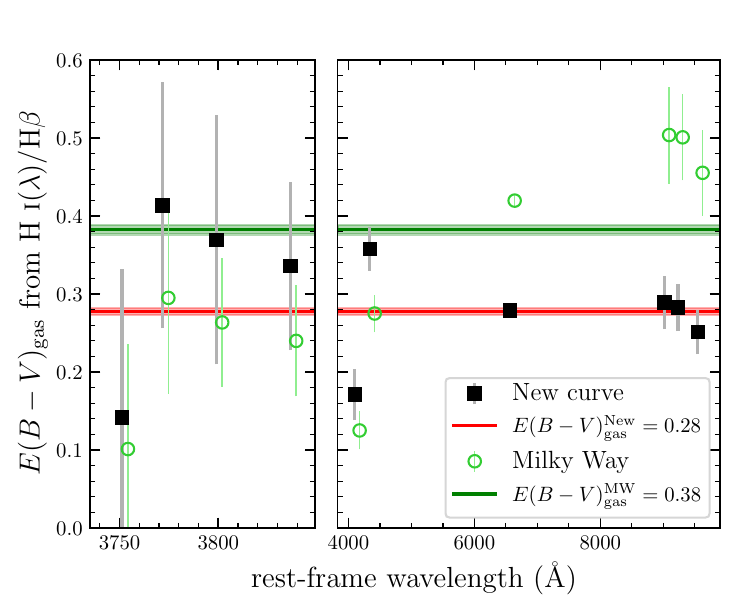}
\caption{\ebvgas\ derived from the measured flux ratio of H\one\ lines relative to H$\beta$ as a function of the rest-frame
 wavelength of each H\one\ transition.
Black squares show the resulting values when the newly derived attenuation curve of GOODSN-17940 is used, while green
 circles display the results when the \citet{car1989} Milky Way curve is assumed.
The \ebvgas\ values calculated by simultaneously fitting all H\one\ lines is shown by the horizontal lines.
}\label{fig:ebvgas}
\end{figure}

Figure~\ref{fig:ebvgas} compares \ebvgas\ derived from different H\one\ lines using the flux ratio H\one($\lambda$)/H$\beta$,
 assuming both the new attenuation curve (black points) and the \citet{car1989} Milky Way curve (green points).
Note that the SMC and \citet{cal2000} curves are very similar in shape to that of the Milky Way at optical wavelengths, and would yield similar results.
This exercise makes it clear that the Milky Way curve is not appropriate for GOODSN-17940 as there is a clear systematic
 trend in \ebvgas\ derived from different H\one\ lines, with an overall increase of the inferred \ebvgas\ as the wavelength of
 the numerator H\one\ line increases.
When using the Milky Way curve, the red lines (H$\alpha$ and Paschen series) yield $E(B-V)_\mathrm{gas}=0.4-0.6$ while lines
 bluer than H$\beta$ imply $E(B-V)_\mathrm{gas}=0.1-0.3$.
This trend has implications for spectroscopic observations of early galaxies at different redshifts with {\it JWST}.
The broad assumption of a Milky Way-like nebular attenuation curve may bias \ebvgas\ inferences above and below $z\sim6.5$.
At $z>6.5$, NIRSpec only covers H$\beta$ and bluer Balmer lines, with H$\gamma$/H$\beta$ commonly used to infer \ebvgas.
At $z<6.5$, H$\alpha$ is accessible with NIRSpec and the traditional H$\alpha$/H$\beta$ Balmer decrement is most often used.
There is no systematic trend in \ebvgas\ when using bluer or redder H\one\ lines with the new curve.

Assuming the Milky Way curve also leads to a higher overall \ebvgas\ than when adopting the new attenuation curve if all available H\one\ lines are used.
The horizontal lines indicate the value of \ebvgas\ obtained when fitting to all 12 H\one\ lines simultaneously,
 yielding $E(B-V)_\mathrm{gas}^\mathrm{New}=0.278\pm0.004$ and $E(B-V)_\mathrm{gas}^\mathrm{MW}=0.382\pm0.006$.
The small errors on these derived reddening values demonstrate that, with deep {\it JWST}/NIRSpec data, the statistical
 uncertainty on dust reddening and attenuation corrections is much smaller than the systematic uncertainty associated
 with the choice of dust curve, such that the error budget is dominated by galaxy-to-galaxy variations.
It is imperative to derive nebular attenuation curves for a larger sample of individual high-redshift galaxies in order to
 understand both the shape of the average curve and the magnitude of variations about this mean.
The full AURORA survey sample will allow for the derivation of nebular attenuation curves for dozens of individual objects
 at $z\gtrsim2$ (Reddy et al., in prep.).

\subsection{Implications for Dust Properties and Geometry}

The dust attenuation curve arises from a combination of the dust grain size, grain chemical composition, and
 the geometric distribution of dust around emitting sources and associated radiative transfer effects.
The significantly different shape of our newly derived attenuation curve relative to the Milky Way, SMC,
 and \citet{cal2000} curves suggests a difference in the dust grain physical properties, dust distribution, or both
 in GOODSN-17940 relative to what is common in local star-forming galaxies.
While we cannot distinguish between the grain property and dust geometry scenarios with the current data,
 we can rule out the case of a foreground dust screen with unity covering fraction if the grain size and composition
 matched that present in the SMC or Milky Way ISM.
Obtaining rest-frame far-infrared observations of GOODSN-17940 to measure its position in the infrared excess (IRX) vs.\ $\beta$
 plane could disentangle geometry from grain property effects \citep[e.g.,][]{pop2017,nar2018}.

We also ruled out the presence of a significant 2175~\AA\ bump in the GOODSN-17940 nebular attenuation curve.
The 2175~\AA\ bump strength in stellar attenuation curves
 has been found to increase with increasing metallicity and decreasing specific SFR
 \citep[e.g.,][]{kri2013,sal2018,shi2020,kas2021}.
High specific SFR galaxies are found to have shallower UV stellar attenuation curves and weaker bumps \citep[e.g.,][]{kri2013},
 which qualitatively agrees with our results for the GOODSN-17940 nebular attenuation curve since its high emission-line
 equivalent widths imply a high specific SFR.
However, a UV bump has been found in the stellar attenuation curves of high-redshift galaxies up to $z\sim7$, implying
 that the more extreme conditions in early galaxies do not preclude its presence \citep{wit2023,mar2023}.

The strength of the 2175~\AA\ bump is thought to be driven by the presence
 of polycyclic aromatic hydrocarbon (PAH) molecules \citep{job1992,li2001,shi2022}, such that processes that
 inhibit the production of PAHs or destroy them reduce the bump strength.
The lack of a significant UV bump in the GOODSN-17940 nebular attenuation curve could be due to
 efficient destruction of PAHs by an intense UV radiation field or the lack of enrichment by
 asymptotic giant branch (AGB) stars that are thought to form PAHs in their carbon-rich winds \citep{shi2020}.
While both scenarios are plausible for a young starburst galaxy like GOODSN-17940, the ultradeep AURORA observations
 can provide a more conclusive answer by enabling direct-method measurements of the C/O and N/O abundance ratios to
 constrain the degree of AGB enrichment and robust constraints on the ionizing radiation field
 (Sanders et al., in prep.).
The 3.3~$\mu$m PAH emission feature is also accessible with MIRI ($\lambda_\mathrm{obs}=17.8~\mu$m at $z=4.41$)
 to directly test whether PAH molecules are present.

\subsection{Comparison to the \citet{red2020} $z\sim2$ Nebular Attenuation Curve}

\citet{red2020} performed the only other direct derivation of the nebular attenuation curve from H\one\ emission lines at high redshifts
 using stacked Keck/MOSFIRE spectra of a representative sample of $\sim500$ star-forming galaxies at $z-1.4-2.6$.
This work was based on 5 H\one\ emission
 lines spanning $4000~\mathrm{\AA}-6600~\mathrm{\AA}$ (H$\epsilon$, H$\delta$, H$\gamma$, H$\beta$, and H$\alpha$).
These authors found an average $z\sim2$ curve that was consistent with the \citet{car1989} Milky Way curve over that wavelength range.
We have repeated our analysis of GOODSN-17940, limiting the lines used to derive the
 attenuation curve to those available in \citet{red2020} (H$\alpha$$-$H$\epsilon$).
When normalizing all curves at 6565~\AA, the resulting GOODSN-17940 attenuation curve agrees within 1$\sigma$ with our curve derived using the
 full suite of H\one\ lines and is inconsistent with the \citet{red2020} average $z\sim2$ curve, displaying a difference of $\Delta k=0.45\pm0.15$ at 5,000~\AA.

A comparison between \citet{red2020} and our analysis demonstrates the ways in which deep {\it JWST}/NIRSpec data is ushering in an era of precision
 reddening and attenuation corrections at high redshifts.
First, the depth and wide wavelength coverage of the AURORA spectra enable the use of more than twice as many H\one\ transitions,
 yielding more robust constraints translating to a smaller uncertainty on $k(\lambda)$ but also a wider wavelength coverage
 by including Paschen-series lines at longer wavelengths and high-order Balmer series lines up to H12 at 3751~\AA.
The latter is key for ensuring a robust correction for line ratios involving [O\ii]$\lambda$3728, typically the bluest rest-optical
 line detected that is crucial for metallicity and ionization state constraints.
Second, NIRSpec allows us to perform this analysis on individual objects whereas \citet{red2020} had to rely on composite spectra.
We can thus understand both the mean high-redshift nebular attenuation curve and the galaxy-to-galaxy variation by applying this
 analysis to the full AURORA survey sample, which should inform the interpretation of larger emission-line galaxy samples from shallower {\it JWST} programs.
Finally, the coverage into the rest-frame far-ultraviolet for $z\gtrsim4$ objects provided by G140M enables the possibility of constraining
 the UV attenuation curve, important for science cases involving rest-UV emission lines such as inferring C/O from C\iii]$\lambda$1908
 or N/O from N\iii]$\lambda$1750
 that have been proposed as a tracers of galaxy formation timescale and exotic enrichment pathways at high redshift
 \citep[e.g.,][]{are2022,jon2023,cam2023,deu2023,sen2024}.

\subsection{Implications for Derived SFRs and Physical Properties}

Star-formation rates derived from dust-corrected H$\alpha$ or other Balmer emission lines are sensitive to
 the assumed nebular dust curve.
It is typical to assume that a single curve applies to all galaxies in the sample when analyzing H$\alpha$-based SFRs
 \citep[e.g.,][]{kas2013,shi2015,shi2016,cla2024}.
For GOODSN-17940, we derive SFR(H$\alpha)>91~\mathrm{M}_\odot~\mathrm{yr}^{-1}$ using the new attenuation curve and $E(B-V)_\mathrm{gas}^\mathrm{New}$
 with the low-metallicity conversion factor of \citet{sha2023},
 a lower limit since we can only set a lower limit to the normalization $R_V>3.50$.
Using the Milky Way curve with $E(B-V)_\mathrm{gas}^\mathrm{MW}$ yields SFR(H$\alpha)=118\pm2~\mathrm{M}_\odot~\mathrm{yr}^{-1}$.
These two SFR values are consistent since the one derived with the new curve is a lower limit.
The Milky Way curve has a lower $k(\mathrm{H}\alpha)$ but yields a higher \ebvgas\ value, while the new curve has a higher $k(\mathrm{H}\alpha)$ but
 yields a lower \ebvgas\ value.
The two curves thus yield consistently high SFRs through a different combination of factors.
It is possible that a significant part of the scatter in the star-forming main sequence based on SFR(H$\alpha$)
 may be caused by the oversimplified assumption of a single dust curve (and $R_V$) for all targets.
It is important to understand such variation and its dependence on galaxy properties
 in studies quantifying the burstiness of star-formation histories by comparing SFR
 derived using different tracers including H$\alpha$ \citep[e.g.,][]{wei2012,guo2016,ema2019,rez2023,asa2024,cla2024}.
However, our results are only for one galaxy with extreme emission-line equivalent widths.
Analysis of the full AURORA sample will provide a more rigorous determination of the scatter in the nebular attenuation curve among
 more representative targets near Cosmic noon.

It is also of note that SFR(H$\alpha$) is $\gtrsim2$ times larger than SFR(SED) when using the SMC curve (Sec.~\ref{sec:sed}).
Since SFR(SED) is primarily sensitive to the UV photometry, this difference is likely driven by the steep UV slope of the SMC curve
 that results in a smaller amount of absolute attenuation when matching the rest-UV slope in the photometry.
When adopting the \citet{cal2000} curve for SED fitting, which has a far-UV slope much closer to what we derived for GOODSN-17940,
 SFR(SED) is in much better agreement with SFR(H$\alpha$).
While a number of works have suggested that an SMC-like curve is appropriate for young high-redshift galaxies on average
 \citep[e.g.,][]{red2018,red2023,shi2020}, our results demonstrate that this assumption is not universally applicable to individual
 systems and can bias the comparison of SFRs derived from different indicators such as H$\alpha$ and UV luminosity,
 or full SED fitting.

The nebular attenuation curve is also of key importance for physical properties derived from line ratios that are
 widely spaced in wavelength, including several common indicators of gas-phase metallicity and ionization parameter
 such as R23=([O\iii]$\lambda$4960,5008+[O\ii]$\lambda$3728)/H$\beta$, O32=[O\iii]$\lambda$5008/[O\ii]$\lambda$3728,
 and N2O2= [N\ii]$\lambda$6585/[O\ii]$\lambda$3728.
The reddening correction of line ratios is sensitive to the shape rather than the normalization of the dust curve.
The reddening correction applied to O32 and N2O2 is $\approx25\%$ (0.1~dex) larger when using the Milky curve and $E(B-V)_\mathrm{gas}^\mathrm{MW}$
 relative to adopting the new curve and $E(B-V)_\mathrm{gas}^\mathrm{New}$.
Given the quality of spectra now available with deep {\it JWST}/NIRSpec observations, 0.1~dex systematic uncertainties on strong-line ratios are now much
 larger than the measurement uncertainties for many high-redshift sources.
{\it JWST} has also enabled robust $T_e$-based direct-method metallicities for an ever-increasing number of high-redshift galaxies
 \citep[e.g.,][]{tru2022,are2022,cur2023,nak2023,san2024,las2024,rog2024,wel2024a,wel2024b}.
The temperature-sensitive auroral-to-strong line ratios (e.g., [O\iii]$\lambda$4364/$\lambda$5008; [O\ii]$\lambda\lambda$7322,7332/[O\ii]$\lambda$3728)
 used to derive $T_e$ require a reddening correction.
When using the Milky Way curve instead of the new curve for GOODSN-17940, the [O\ii] temperature is 9\%\ smaller and the derived O$^+$/H ratio is 0.2~dex larger (Sanders et al., in prep.),
 demonstrating that a correct choice of dust curve is required to achieve precise chemical abundance constraints.
The ability to derive attenuation curves for individual objects provided by deep NIRSpec observations
 will thus lead to a significant reduction in the uncertainties on inferred ISM physical properties from emission-line ratios.

\subsection{Potential Systematic Uncertainties}

Given the complexity of the flux calibration and slit loss correction for {\it JWST}/NIRSpec MSA data,
 it is of interest to assess whether uncertainties in these processes could drive the deviations we see
 from the canonical dust curves.
Our analysis makes use of data and line ratios across all three medium-resolution NIRSpec gratings (G140M, G235M, and G395M),
 such that the relative flux calibration between gratings is one concern.
We can assess the relative calibration using overlapping wavelength coverage between neighboring gratings.
For both G140M and G395M, the integrated flux density in the $>2000$~\AA\ of overlap with G235M agrees to better than 10\%, indicating that
 systematics associated with the relative flux calibration are $<10\%$.
For H$\alpha$ (in G395M) to yield \ebvgas\ consistent with the H\one\ lines bluer than H$\beta$ (in G235M) when assuming the Milky Way curve,
 the H$\alpha$ flux would need to decrease by $20-30\%$.
However, shifting the flux level of G395M relative to G235M cannot account for the observed trend in H\one\ line ratios as systematic differences
 in the inferred \ebvgas\ remain even when only using ratios of lines in the same grating.
The H\one\ lines from H$\beta$ bluewards are all in G235M, and the \ebvgas\ values derived from these lines using the Milky Way curve are $0.1-0.3$.
If we instead use the ratio of Pa8, Pa9, and Pa10 to H$\alpha$ (all in G395M) to calculate \ebvgas, the resulting values are $\approx0.6$
 using the Milky Way curve.
Uncertainty in the NIRSpec flux calibration thus cannot account for the trends we see.

Likewise, an inaccurate slit loss correction also cannot cause our measurements to deviate from expectations based on a Milky Way-like curve.
The wavelength-dependent slit losses are driven by the change in size of the {\it JWST}/NIRSpec PSF as a function of wavelength relative to the fixed size of the microshutters.
Since the PSF size increases monotonically with wavelength, slit loss effects are a monotonic function of wavelength.
Making both the H\one\ lines bluewards of H$\beta$ and the red lines (H$\alpha$ and Paschen series) simultaneously agree with
 expectations from the Milky Way curve would require both the high-order Balmer lines (H$\gamma$ to H12) and the H$\alpha$+Pa lines to decrease in flux relative to H$\beta$.
This hypothetical shift is not monotonic with wavelength and thus cannot be driven by PSF-related slit loss effects.

Measured H\one\ line fluxes can be impacted by underlying stellar absorption such that an incorrect
 stellar population model may bias the H\one\ line ratios.
Due to the extremely large equivalent widths of the emission lines of GOODSN-17940, seemingly driven
 by a very young age and high specific SFR, varying the assumed continuum model during emission line fitting
 does not significantly change the derived dust curve.
Changing the continuum from the fiducial SMC model to the Calzetti model or a simple linear continuum that provides no stellar
 absorption correction does not significantly impact our results.
We conclude that the distinct shape of the GOODSN-17940 nebular attenuation curve is not caused by systematics associated
 with the flux calibration, slit loss correction, or stellar absorption correction.

In our derivation of the UV attenuation curve, we assumed that the dust reddening and attenuation affecting the nebular gas is
 the same as that affecting the stars dominating the UV continuum due to the extremely young age ($\sim$5~Myr) of GOODSN-17940.
If we instead assume differential reddening such that the stars experience less reddening than the nebular gas, as found in
 local starbursts and high-redshift star-forming galaxies with larger average ages than GOODSN-17940 \citep[e.g.,][]{cal1997,cal2000,red2015},
 then the inferred UV attenuation curve would steepen relative to the fiducial curve presented here.
A UV curve as steep as the SMC curve could only be accommodated if the nebular gas experiences more than three times the amount of reddening as the stars.
The 2175~\AA\ bump remains strongly ruled out even in this case.

\section{Summary and Conclusions}\label{sec:conclusions}

We have used ultradeep medium-resolution $1-5$~$\mu$m {\it JWST}/NIRSpec observations from the AURORA survey to
 derive the nebular dust attenuation curve for the galaxy GOODSN-17940 at $z=4.41$ over rest-frame ultraviolet to near-infrared wavelengths.
We utilized measured flux ratios of 12 H\one\ Balmer and Paschen series recombination lines
 to constrain the shape of the attenuation curve over $3750-9550$~\AA.
We then leveraged a high-S/N spectroscopic detection of the rest-frame ultraviolet continuum,
 comparing the observed continuum shape to an assumed intrinsic UV slope to constrain the shape of the UV attenuation curve
 down to 1400~\AA.
Since the age of the stellar population of GOODSN-17940 is $<10$~Myr based on the measured equivalent widths of H$\alpha$ and H$\beta$
 and consistent SED fitting results, this UV constraint based on the combined stellar and nebular continuum effectively probes the nebular attenuation
 curve since the young stars and ionized gas should be spatially coincident.
The combined nebular attenuation curve spanning $1400-9550$~\AA, presented in Figure~\ref{fig:totcurve} and equation~\ref{eq:ktot},
 displays a shape that is significantly different than the most commonly assumed dust curves in extragalactic studies.
It is steeper than the Milky Way, SMC, and \citet{cal2000} curves on average over optical wavelengths,
 but shallower in the ultraviolet and lacks a 2175~\AA\ bump.
 
The strong deviation of the measured attenuation curve from the commonly assumed ones highlights the degree to which
 dust curves can vary among individual high-redshift galaxies and carries important implications for the measurement of physical properties
 that require corrections for dust.
We demonstrated that the inferred \ebvgas\ is systematically biased when assuming a Milky Way-like curve, with H$\alpha$ and Paschen-series lines
 yielding \ebvgas\ that is $0.2-0.3$~mag higher than when using the blue Balmer lines (e.g., H$\gamma$/H$\beta$).
This result carries implications for {\it JWST}/NIRSpec observations of galaxies at different redshifts, where coverage is only possible for
 H$\beta$ and bluer H\one\ lines at $z\gtrsim6.5$ while H$\alpha$ is accessible at lower redshifts.
Significant galaxy-to-galaxy variation of the nebular attenuation curve may account for part of the observed scatter in the star-forming main sequence
 when using H$\alpha$-based SFRs and offsets between SFRs based on H$\alpha$ and UV luminosity that have been interpreted as evidence for bursty star formation.
It is imperative to characterize the magnitude of this variation and its dependence on galaxy properties to properly interpret these trends.
The use of appropriate nebular attenuation curves is also needed to improve the accuracy of physical properties derived from emission line ratios
 including metallicity and ionization parameter.

The robust characterization of the ultraviolet-to-near-infrared nebular attenuation curve for a $z=4.41$ star-forming galaxy
 presented here demonstrates the diagnostic power of deep and continuous $1-5~\mu$m spectroscopy with {\it JWST}/NIRSpec.
While the new dust curve will improve the inference of physical properties for GOODSN-17940, this target is
 ultimately only a single galaxy.
The SFR of GOODSN-17940 is $\approx20\times$ elevated relative to the $z=4-5$ main sequence at $10^9$~M$_\odot$ \citep[e.g.,][]{spe2014,cla2024}.
Its large emission-line equivalent widths are rare at $z\sim2-4$ \citep{red2018a,tan2019,boy2022},
 but are more typical of UV-bright galaxies at $z>6$ where H$\alpha$ and [O\iii]+H$\beta$ equivalent widths $>1000$~\AA\ are common
 \citep{end2021,end2023,end2024,hei2024}.
This source is thus not representative of typical galaxies at $z\sim4.5$, but instead is an analog to bright
 reionization-epoch sources, implying that the attenuation curves of such galaxies may significantly deviate
from what is commonly assumed.
This work paves the way for similar analyses of larger samples of galaxies including upcoming
 work with the full AURORA survey sample to characterize the variation in the nebular dust curve among individual sources (Reddy et al., in prep.).
Refinements to the accuracy of dust reddening and attenuation corrections made possible with {\it JWST}/NIRSpec will ultimately lead
 to improved constraints on the chemical and ionization state of the ISM and SFRs at high redshifts,
 advancing our knowledge of galaxy formation and evolution in the early Universe.

\clearpage

\begin{acknowledgments}
This work is based on observations made with the NASA/ESA/CSA James Webb Space Telescope. The data were
obtained from the Mikulski Archive for Space Telescopes at
the Space Telescope Science Institute, which is operated by the
Association of Universities for Research in Astronomy, Inc.,
under NASA contract NAS5-03127 for JWST.  The specific observations analyzed can be accessed via \dataset[DOI: 10.17909/hvne-7139]{https://archive.stsci.edu/doi/resolve/resolve.html?doi=10.17909/hvne-7139}.
Some of the data products presented herein were retrieved from the Dawn JWST Archive (DJA). DJA is an initiative of the Cosmic Dawn Center (DAWN), which is funded by the Danish National Research Foundation under grant DNRF140.
We also acknowledge support from NASA grant JWST-GO-01914.
FC acknowledges support from a UKRI Frontier Research Guarantee Grant (PI Cullen; grant reference: EP/X021025/1).
CTD, DJM, RJM, and JSD acknowledge the support of the Science and Technology Facilities Council.
JSD also acknowledges the support of the Royal Society through a Royal Society Research Professorship.
RD acknowledges support from the Wolfson Research Merit Award program of the U.K. Royal Society.
KG acknowledges support from the Australian Research Council Laureate Fellowship FL180100060.
MK acknowledges funding from the Dutch Research Council (NWO) through the award of the Vici grant VI.C.222.047 (project 2010007169).
PO acknowledges the Swiss State Secretariat for Education, Research and Innovation (SERI) under contract number MB22.00072, as well as from the Swiss National Science Foundation (SNSF) through project grant 200020$\_$207349.
AJP was generously supported by a Carnegie Fellowship through the Carnegie Observatories.
RSE acknowledges financial support from the Peter and Patricia Gruber Foundation.
ACC acknowledges support from a UKRI Frontier Research Guarantee Grant [grant reference EP/Y037065/1].
DN was funded by JWST-AR-01883.001.
Finally, we thank members of the JADES team for assistance with target selection in the GOODS-N field.
\end{acknowledgments}

\vspace{5mm}
\facilities{{\it JWST} (NIRSpec, NIRCam)}

\bibliography{dust}{}
\bibliographystyle{aasjournal}

\end{document}